\documentclass[twocolumn,pre,aps,showpacs,showkeys,amsmath]{revtex4}
\usepackage{graphicx}
\usepackage{adjustbox}
\usepackage{multirow}

\begin{document}

\title{Functions of Direct and Indirect Pathways for Action Selection Are Quantitatively Analyzed in A Spiking Neural Network of The Basal Ganglia
}
\author{Sang-Yoon Kim}
\email{sykim@icn.re.kr}
\author{Woochang Lim}
\email{wclim@icn.re.kr}
\affiliation{Institute for Computational Neuroscience and Department of Science Education, Daegu National University of Education, Daegu 42411, Korea}

\begin{abstract}
We are concerned about action selection in the basal ganglia (BG). We quantitatively analyze functions of direct pathway (DP) and indirect pathway (IP) for action selection in a spiking neural network with 3 competing channels. For such quantitative analysis, in each channel, we obtain the competition degree
${\cal C}_d$, given by the ratio of strength of DP (${\cal S}_{DP}$) to strength of IP (${\cal S}_{IP}$) (i.e., ${\cal C}_d = {\cal S}_{DP} / {\cal S}_{IP}$).
Then, a desired action is selected in the channel with the largest ${\cal C}_d$. Desired action selection is made mainly due to strong focused inhibitory projection to the output nucleus, SNr (substantia nigra pars reticulata) via the DP in the corresponding channel. Unlike the case of DP, there are two types of IPs; intra-channel IP and inter-channel IP, due to widespread diffusive excitation from the STN (subthalamic nucleus). The intra-channel IP serves a function of brake to suppress the desired action selection.  In contrast, the inter-channel IP to the SNr in the neighboring channels suppresses competing actions, leading to highlight the desired action selection. In this way, function of the inter-channel IP is opposite to that of the intra-channel IP. However, to the best of our knowledge, no quantitative analysis for such functions of the DP and the two IPs was made. Here, through direct calculations of the DP and the intra- and the inter-channel IP presynaptic currents into the SNr in each channel, we obtain the competition degree of each channel to determine a desired action, and then functions of the DP and the intra- and inter-channel IPs are quantitatively made clear.
\end{abstract}



\maketitle

\section{Introduction}
\label{sec:INT}
The basal ganglia (BG) (i.e., a group of subcortical deep-lying nuclei) in the brain receive cortical inputs from most areas of cortex, and
provide inhibitory output projection to the thalamus/brainstem \cite{Luo,Kandel,Squire,Bear}.
The BG show diverse functions for motor (e.g., initiation and execution of movement) \cite{Luo,Kandel,Squire,Bear} and cognition (e.g., action selection)
\cite{GPR1,GPR2,Hump1,Hump2,Hump3,Man,Spin,Triple}. Dysfunction in the BG is associated with movement disorder (e.g., Parkinson's disease and Huntington's disease) and cognitive disorder such as dementia \cite{Luo,Kandel,Squire,Bear}.

Diverse subjects in the BG were investigated in computational works by employing a variety of neuron models.
We take some examples of BG neuron models;
(i) artificial neuron model of leaky-integrator type \cite{GPR1,GPR2,Hump3},
(ii) leaky integrate-and-fire model \cite{Hump1,Hump2},
(iii) Izhikevich neuron model \cite{SPN1,Str2,CN13,CN15,CN7,SPN2,CN20,Man,CN2,CN3,CN4,CN14,Fount,CN9,Spin, CN5, CN18,CN1},
(iv) adaptive exponential integrate-and-fire model \cite{CN11,CN12,CN19}, and
(v) point neuron function using the rate-coded output activation \cite{CN10,Frank1,Frank2}.
In this paper, we pay attention to action selection performed by the BG in a spiking neural network (SNN) of the BG with 3 laterally interconnected channels; each channel represents an action. We employ the same SNN with the single channel as that considered in our prior works \cite{KimPD,KimHD}.
Each single channel takes excitatory input from most areas of cortex through the input nuclei [striatum and subthalamic nucleus (STN)] and provide
inhibitory output via the output nucleus [substantia nigra pars reticulata (SNr)] to the thalamus/brainstem \cite{Hump1,Man,Spin}.
The striatum (corresponding to the principal input nucleus) is also the primary recipient of dopamine (DA), coming from the substantia nigra pars compacta (SNc). The only primary output neurons in the striatum are just spine projection neurons (SPNs) which comprise up to 95 $\%$ of the whole striatal population \cite{Str1,Str2}. There exist two kinds of SPNs with D1 and D2 receptors for the DA. Firing behaviors of the D1 and D2 SPNs are modulated in a different way by the DA \cite{SPN1,SPN2,Fount}.

In the SNN with a single channel, there are direct pathway (DP) and indirect pathway (IP) \cite{DIP1,DIP2,DIP3,DIP4}.
Through the DP, focused inhibition from the D1 SPNs is provided onto the output nucleus, SNr, which leads to decrease in the firing activity of the SNr.
Thus, the thalamus becomes disinhibited, resulting in a desired action selection (i.e., ``Go'' behavior).
On the other hand, D2 SPNs are indirectly connected to the SNr through the IP, crossing the intermediate GP (globus pallidus) and the STN.
In this case, the IP serves a function of brake to suppress the desired action selection (i.e., ``No-Go'' behavior), because
the firing activity of the SNr becomes increased mainly due to excitation from the STN.
In the above sense, the DP and the IP are also called the ``Go'' and ``No-Go'' pathways, respectively \cite{Frank1,Frank2,Go1,Go2}.
A variety of functions of the BG may be done via harmony between the ``Go'' DP and the ``No-Go'' IP, and such harmony is regulated by the DA level \cite{Luo,Kandel,Squire,Bear}. Recently, we introduced the competition degree ${\cal C}_d$ between DP and IP, given by the ratio of strength of DP (${\cal S}_{DP}$) to strength of IP (${\cal S}_{IP}$) (i.e., ${\cal C}_d = {\cal S}_{DP} / {\cal S}_{IP}$) \cite{KimPD,KimHD}, and quantified their harmony.

In contrast to the case of single channel, in the SNN with 3 competing channels, diffusive excitation from the STN is given to the SNr and the GP in all the 3 channels \cite{Parent1,Parent2,Parent3}, which results in increase in the spiking activity of the SNr in all the channels. Through these widespread diffusive divergent excitation from the STN, inter-channel connections are made. Thus, there appear two kinds of IPs, intra-channel IP and inter-channel IP.
As in the case of single channel, a desired action is selected through focused inhibition from the D1 SPNs via the Go DP in a channel (off-center effect).
Also, the intra-channel IP is just the above No-Go IP, serving a function of brake to suppress the desired action selection in the corresponding channel.
In contrast, the inter-channel IP to the SNr in the neighboring channels serves a function of suppressing competing actions, which results in
highlighting the desired action selection (on-surround effect, causing contrast enhancement) \cite{Mink1,Mink2,Nambu,Triple,GPR1,GPR2,Hump1,Hump2,Hump3}.
Due to the function of the inter-channel IP, no interference between the desired action and the competing actions occurs.
In this way, functions of the intra-channel and the inter-channel IPs are opposite.

But, to the best of our knowledge, no quantitative analysis for the functions of the DP and the two intra- and inter-channel IPs was made.
In this paper, we make quantitative analysis of functions of DP and IP for action selection, based on the recently-introduced competition degree ${\cal C}_d$ (making characterization of competitive harmony between DP and IP) \cite{KimPD,KimHD}. Through computations of the DP and the intra- and inter-channel IP synaptic currents into the SNr in each channel, we obtain the competition degree ${\cal C}_d$ of each channel to determine a desired action. Thus, functions of the DP and the intra- and inter-channel IPs (causing the off-center and the on-surround effect) are quantitatively made clear.

This paper is organized as follows. In Sec.~\ref{sec:DGN}, we make brief description of the SNN with 3 laterally interconnected channels.
In the Supplementary Information (SI), brief description on the SNN with a single channel is given.
Then, in the main Sec.~\ref{sec:QA}, we quantify the functions of the DP and the intra- and inter-channel IPs for action selection, based on the competition degree,
by calculating the DP and the intra- and inter-channel IP synaptic currents into the SNr in each channel.
Finally, summary and discussion are given in Sec.~\ref{sec:SUM}.

\begin{figure}
\includegraphics[width=0.9\columnwidth]{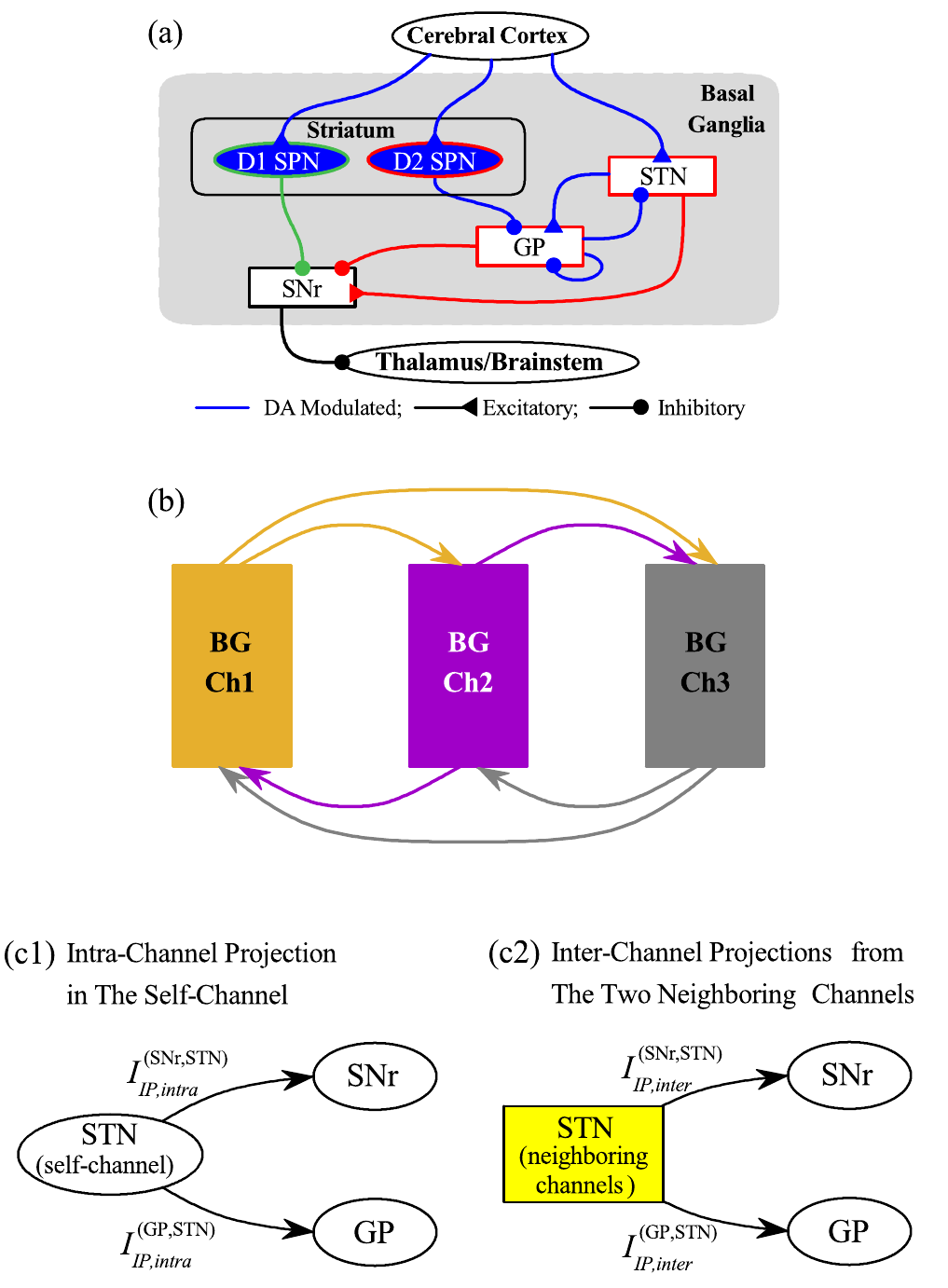}
\caption{Spiking neural network (SNN) for the basal ganglia (BG). (a) Box diagram for the SNN with a single channel. Lines with triangles and circles represent the excitatory and inhibitory connections, respectively. Blue colored cells and lines denote BG cells and synaptic connections affected by the dopamine. Striatum and subthalamic nucleus (STN) receive the cortical input. In the striatum, there are two types of inhibitory spine projection neurons (SPNs); D1 (D2) SPNs with the D1 (D2) receptors. Inhibitory projection from the D1 SPNs to the substantia nigra pars reticulata (SNr) via the direct pathway (DP) is denoted by green line. D2 SPNs are indirectly linked to the SNr via the indirect pathway (IP) represented by red lines; the IP crosses the globus pallidus (GP) and the STN. These DP and IP compete and control inhibitory output from the SNr to the thalamus/brainstem. (b) Box diagram for the SNN with three laterally interconnected channels.
The channels 1, 2, and 3 are denoted in orange, purple, and gray color, respectively. There are inter-channel connections from neighboring channels. Diagrams for (c1) the intra-channel projection from STN to SNr $I_{IP,intra}^{({\rm SNr,STN})}$ and GP $I_{IP,intra}^{({\rm GP,STN})}$ in the same self-channel and (c2) the inter-channel projections from STN in the two neighboring channels (denoted in yellow color) to SNr $I_{IP,inter}^{({\rm SNr,STN})}$ and GP $I_{IP,inter}^{({\rm GP,STN})}$.
}
\label{fig:BGN}
\end{figure}

\section{Spiking Neural Network with Three Competing Channels}
\label{sec:DGN}
We consider an SNN with 3 laterally interconnected channels for action selection in the BG \cite{Hump1,Spin}.
Here, each single channel is the same as that considered in our prior works \cite{KimPD,KimHD}.
The SNN with a single channel is founded on anatomical and physiological data obtained in rat-based works.
The framework (e.g., number of BG neurons and synaptic connection probabilities) is based on the anatomical works \cite{Ana1,Ana2,Ana3,Ana4}.
The intrinsic parameter values of single BG neurons are based on the physiological properties of the BG neurons \cite{Phys1,Phys2,Phys3,Phys4,Phys5,Phys6,Phys7,Phys8,Phys9,Phys10,Phys11}, and the synaptic parameters (related to synaptic currents) are also based on the physiological works \cite{Phys12,Phys13,Phys14,Phys15,Phys16,Phys17,Phys18,Phys19,Phys20}.
Here, we employ rat-brain terminology.

Figure \ref{fig:BGN}(a) shows a box diagram for the SNN with a single channel, composed of D1/D2 SPNs, STN neurons, GP neurons, and SNr neurons. Blue colored cells and lines represent BG cells and synaptic connections, affected by the DA, respectively. Both striatum and STN receive cortical inputs from most regions of the cortex. We model cortical inputs in terms of 1,000 independent Poisson spike trains with the same firing rate $f_{\rm Ctx}$. There are two pathways, DP (green) and IP (red). Inhibitory projection from the D1 SPNs to the output nucleus SNr is provided via the DP. On the other hand, D2 SPNs are indirectly linked to the SNr via the IP, crossing the GP and the STN. Inhibitory output from the SNr to the thalamus/brainstem is controlled via competitive harmony between DP and IP \cite{KimPD}. In the SI, brief description on the SNN with a single channel is given; for details, refer to Sec.~II in \cite{KimPD}.

Figure \ref{fig:BGN}(b) shows a box diagram of the SNN with 3 competing channels; channels 1, 2, and 3 are represented in orange, purple, and gray color, respectively. Here, each channel represents an action \cite{GPR1,Hump1}. We note that inter-channel connections are made via widespread diffusive excitation from the STN in a channel to the target nuclei, SNr and GP, in all the 3 channels \cite{Parent1,Parent2,Parent3}. Thus, there appear one intra-channel interaction in the
same self-channel and two inter-channel interactions from different neighboring channels. Figure \ref{fig:BGN}(c1) shows the intra-channel IP synaptic currents from STN to SNr and GP in the same self-channel, $I_{IP,intra}^{({\rm SNr,STN})}$ and $I_{IP,intra}^{({\rm GP,STN})}$, respectively. In contrast, the inter-channel IP synaptic currents from STN neurons in the two neighboring channels (represented in yellow color) to SNr and GP, $I_{IP,inter}^{({\rm SNr,STN})}$ and $I_{IP,inter}^{({\rm GP,STN})}$, are shown in Fig.~\ref{fig:BGN}(c2). Thus, to the target neurons, SNr and GP, in a channel, there are one intra-channel IP synaptic current from the source STN in the self-channel and two inter-channel IP synaptic currents from the source STN in the two neighboring channels.
For example, we can consider the target nuclei, SNr and GP, in the channel 1. In this case, the intra-channel IP synaptic currents into the target nuclei come from the source STN in the same channel 1, while the two inter-channel IP synaptic currents into the target nuclei in the channel 1 come from the source STN in the neighboring channels 2 and 3. The multi-channel (MCh) connection probability $p_{(c,MCh)}^{(T,STN)}$ from the source STN to the target (SNr, GP) in both cases of intra- and inter-channel interactions is given by $p_{(c,MCh)}^{(T,STN)} = p_c^{(T,STN)}/N_C$; $p_c^{(T,STN)}$ is the connection probability for the case of single channel and $N_C$ is the number of channels \cite{Hump1}. Then, the number of afferent synapses into the target neurons becomes constant, independently of $N_C$. In the present work, $N_C=3.$

A desired action may be selected via focused inhibition from the D1 SPNs through the DP in a channel (off-center effect). In this case, the intra-channel IP serves a function of brake to suppress the desired action selection in the corresponding self-channel.
On the other hand, the inter-channel IP to the SNr in the neighboring channels serves a function of suppressing competing actions \cite{Mink1,Mink2,Nambu,Triple,GPR1,GPR2,Hump1,Hump2,Hump3}. Because of the function of the inter-channel IP, the on-surround effect, causing contrast enhancement, occurs, leading to spotlight the desired action selection. Consequently, there occurs no interference between the desired action and the competing actions.

\section{Quantitative Analysis of Functions of The DP and The Intra- and Inter-channel IP for Action Selection}
\label{sec:QA}
In this section, we consider the SNN with 3 laterally interconnected channels for action selection in Fig.~\ref{fig:BGN}(b).
Here, we also consider a healthy state for a normal DA level ($\phi =0.3$), and make quantitative analysis of functions of the DP and the two intra- and inter-channel IPs by employing the competition degree ${\cal C}_d$ (characterizing competitive harmony between DP and IP) \cite{KimPD,KimHD}.
Through calculations of the DP and the intra- and inter-channel IP synaptic currents into the SNr in each channel, we get the competition degree ${\cal C}_d$ of each channel to determine a desired action. Through such process, we quantitatively make clear functions of the DP and the intra- and inter-channel IPs [causing focused selection (off-center effect) and diffusive inhibition of competing actions (on-surround effect)].

\begin{figure}
\includegraphics[width=0.8\columnwidth]{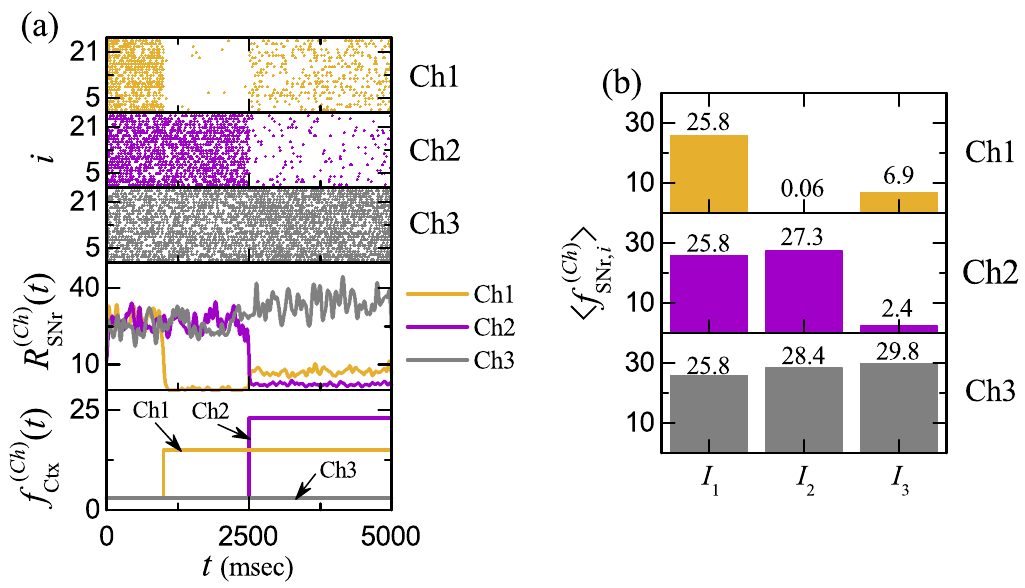}
\caption{Population and individual firing behaviors of SNr neurons in each channel. (a) Raster plots of spikes of SNr neurons, IPSRs (instantaneous population spike rates) $R_{\rm SNr}^{(Ch)}$ (t) of SNr neurons, and frequencies $f_{\rm Ctx}^{(Ch)}$ (t) of cortical inputs in the three channels [channel 1 (Ch1: orange); channel 2 (Ch2: purple); channel 3 (Ch3: gray)]. (b) Histograms of population-averaged mean firing rates $\langle f_{{\rm SNr},i}^{(Ch)} \rangle$ of SNr neurons in the 3 channels for each time intervals $I_1$, $I_2$, and $I_3$.}
\label{fig:PIFB}
\end{figure}

\subsection{Population and Individual Firing Behaviors in The Output Nucleus, SNr}
\label{subsec:PIFB}
Figure \ref{fig:PIFB} shows population and individual firing behaviors in the 3 channels; channel 1 (Ch1; orange), channel 2 (Ch2: purple), and channel 3 (Ch3: gray). In the 1st time interval $I_1 (= [0, 1,000]$ msec), tonic cortical inputs of frequency $f_{\rm Ctx}^{(Ch)}=3$ Hz are provided to all the 3 channels.
From $t=1,000$ msec, the channel 1 receives a cortical input of frequency $f_{\rm Ctx}^{(1)}=15$ Hz, while the channel 2 receives more salient cortical input of frequency $f_{\rm Ctx}^{(2)}=23$ Hz from $t=2,500$ sec. Thus, in the 2nd time interval $I_2$ (= [1,000, 2,500] msec), the channel 1 is in the phasically-active state receiving 15 Hz cortical input, while the other channels 2 and 3 are in the resting default state receiving tonic cortical input (3 Hz). In the last 3rd time interval $I_3$ (= [2,500, 5,000] msec), the channels 1 and 2 receive cortical inputs of frequencies $f_{\rm Ctx}^{(1)}=15$ and $f_{\rm Ctx}^{(2)}=23$ Hz, respectively, while the channel 3 continues to receive tonic cortical input (3 Hz). The frequencies $f_{\rm Ctx}^{(Ch)}(t)$ of cortical inputs in the 3 channels are well shown in the bottom row of Fig.~\ref{fig:PIFB}(a).

The SNr is the output nucleus of the BG, providing inhibitory projection to the thalamus.
Population firing activity of the SNr neurons may be well visualized in the raster plot of spikes which is a collection of spike trains of individual SNr neurons. The top 3 rows of Fig.~\ref{fig:PIFB}(a) show the raster plots of spikes for the 26 SNr neurons in the 3 channels.
In the case of channel 1, spikes appear very sparsely in the time interval $I_2$ (where the channel 1 receives cortical input with $f_{\rm Ctx}^{(1)}=15$ Hz), while in the case of channel 2, spikes occur in a relatively sparse way in the time interval $I_3$ (where the channel 2 receives cortical input with $f_{\rm Ctx}^{(2)}=23$ Hz). Unlike the cases of channels 1 and 2, in the case of channel 3,
spikes tend to appear a little more densely with increasing the time interval due to increased inter-channel IP synaptic currents from the channels 1 and 2.

As a collective quantity showing population behaviors, we use an IPSR (instantaneous population spike rate) which may be obtained from the raster plot of spikes
\cite{W_Review,Sparse1,Sparse2,Sparse3,FSS,SM}. To get the smooth IPSR, we employ the kernel density estimation (kernel smoother) \cite{Kernel}. Each spike in the raster plot is convoluted (or blurred) with a kernel function $K_h(t)$ to get a smooth estimate of IPSR $R_{\rm SNr}^{(Ch)}(t)$ ($Ch=$ 1, 2, and 3):
\begin{equation}
R_{\rm SNr}^{(Ch)}(t) = \frac{1}{N_{\rm SNr}} \sum_{i=1}^{N_{\rm SNr}} \sum_{s=1}^{n_i} K_h (t-t_{s,i}),
\label{eq:IPSR}
\end{equation}
where $N_{\rm SNr}$ $(= 26)$ is the number of the SNr neurons, $t_{s,i}$ is the $s$th spiking time of the $i$th SNr neuron,
$n_i$ is the total number of spikes for the $i$th SNr neuron, and we use a Gaussian kernel function of band width $h$:
\begin{equation}
K_h (t) = \frac{1}{\sqrt{2\pi}h} e^{-t^2 / 2h^2}, ~~~~ -\infty < t < \infty,
\label{eq:Gaussian}
\end{equation}
where the band width $h$ of $K_h(t)$ is 20 msec.

The IPSRs $R_{\rm SNr}^{(Ch)}(t)$ in the 3 channels are also shown in Fig.~\ref{fig:PIFB}(a).
In the case of channel 1 (receiving cortical inputs of $f_{\rm Ctx}^{(1)}=15$ Hz in the time intervals $I_2$ and $I_3$), $R_{\rm SNr}^{(1)}(t)$ drops very rapidly in the time interval $I_2$, and then it increases a little in the time interval $I_3$ (due to the inter-channel IP synaptic current from the channel 2). On the other hand, in the channel 2 (receiving cortical input of $f_{\rm Ctx}^{(2)}=23$ Hz during the time interval $I_3$), $R_{\rm SNr}^{(2)}(t)$ decreases rapidly in the time interval $I_3$. For the channel 3 (receiving tonic cortical inputs of $f_{\rm Ctx}^{(3)}=3$ Hz), $R_{\rm SNr}^{(3)}(t)$ tends to increase slowly with increasing the time interval (because of increase in the inter-channel IP synaptic currents from the channels 1 and 2).

\begin{figure}
\includegraphics[width=0.8\columnwidth]{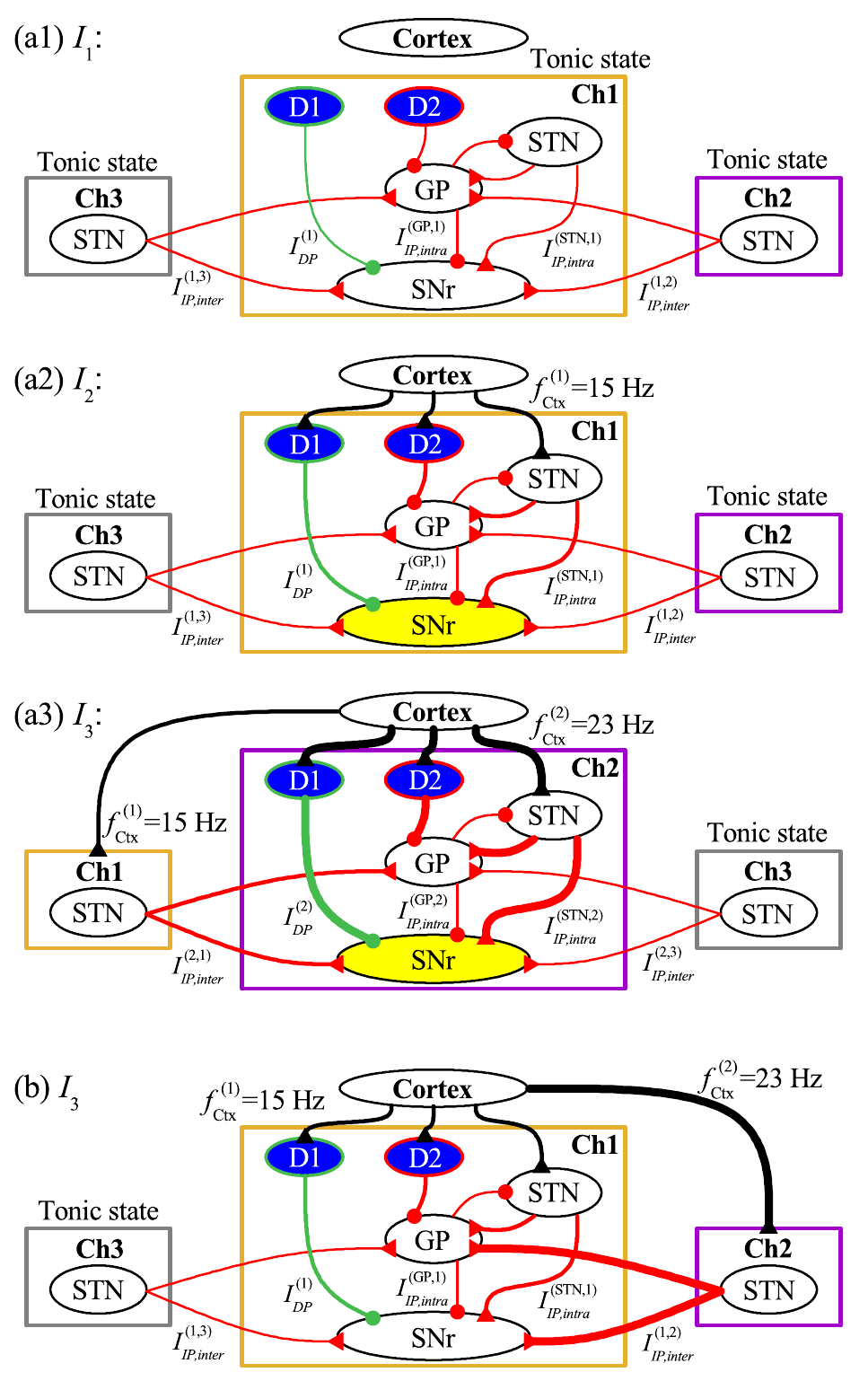}
\caption{Action selection and switching. Diagrams of change in the DP and the intra- and inter-channel IP synaptic currents in the time intervals (a1) $I_1$, (a2) $I_2$, and (a3) $I_3$. DP and IP are represented in green and red colors, respectively; cortical inputs are denoted in black color.
 The width of a line denotes strength of synaptic current. No action selection in $I_1$. Action selection of the channel 1 in $I_2$ (off-center effect due to focused strong DP synaptic current). Action switching to the channel 2 from the channel 1 in $I_3$. Ch1 (channel 1: orange), Ch2 (channel 2: purple), and Ch3 (channel 3: gray). Action-selected channels are shaded in yellow color. (b) Diagram for action deselection of the channel 1 in the time interval $I_3$ (on-surround effect due to strong inter-channel synaptic current from the channel 2).
}
\label{fig:ASS}
\end{figure}

We also study the (interval-averaged) mean firing rates (MFRs) $f_{{\rm SNr},i}^{(Ch)}$ ($i=1, 2, \cdots, N_{\rm SNr}$; $Ch=$ 1, 2, and 3) of individual SNr neurons. For each individual SNr neuron, we get its interval-averaged MFR in a time interval $I$ by dividing the number of spikes $N_s$ occurring during $I$ with the time interval $I$;
\begin{equation}
f_{{\rm SNr},i}^{(Ch)} = {\frac {N_s} {I}}.
\label{eq:MFR}
\end{equation}
Then, we obtain their population-averaged MFR $\langle f_{{\rm SNr},i}^{(Ch)} \rangle$;
\begin{equation}
\langle f_{{\rm SNr},i}^{(Ch)} \rangle  = {\frac {1} {N_{\rm SNr}}} \sum_{i=1}^{N_{\rm SNr}} f_{{\rm SNr},i}^{(Ch)}.
\label{eq:PMFR}
\end{equation}

Firing activity of the SNr (i.e., output nucleus of the BG) is well characterized in terms of their population-averaged MFR $\langle f_i^{({\rm SNr})} \rangle$.
When $\langle f_i^{({\rm SNr})} \rangle$ is high (low), the BG gate to the thalamus becomes locked (opened), leading to inhibition (disinhition) of the thalamus. In this way, the population-averaged MFR of the SNr, $\langle f_i^{({\rm SNr})} \rangle$, is a good indicator for the output activity of the BG,
and hence it could also be used to determine a desired action selection \cite{Hump1, Spin}.
Figure \ref{fig:PIFB}(b) shows population-averaged MFR $\langle f_{{\rm SNr},i}^{(Ch)} \rangle$ in each time interval $I_1$, $I_2$, and $I_3$ for the channel 1 (Ch1; orange), channel 2 (Ch2: purple), and channel 3 (Ch3: gray). We also note that the population-averaged MFR $\langle f_{{\rm SNr},i}^{(Ch)} \rangle$ in each time interval $I_j$ ($j=$ 1, 2, and 3) is just the same as the interval-averaged IPSR, $\overline{ R_{\rm SNr}^{(Ch)}(t)}$ (averaged one in each interval); the overline represents the time-averaging.

In the 1st time interval $I_1$, the population-averaged MFRs of the SNr neurons in the 3 channels (receiving the cortical inputs with $f_{\rm Ctx}^{(Ch)} = 3$ Hz) are the same (i.e., $\langle f_{{\rm SNr},i}^{(1)} \rangle = \langle f_{{\rm SNr},i}^{(2)} \rangle = \langle f_{{\rm SNr},i}^{(3)} \rangle = 25.8$ Hz). Due to strong firing activity of the SNr neurons, the thalamus is inhibited (i.e., the BG gate to the thalamus becomes locked), leading to no action selection. But, in the 2nd time interval $I_2$, $\langle f_{{\rm SNr},i}^{(1)} \rangle$ in the 1st channel (receiving cortical input with $f_{\rm Ctx}^{(1)} = 15$ Hz) becomes so much decreased to 0.06 Hz due to focused inhibition from D1 SPNs via DP, leading to disinhibition of the thalamus (causing action selection). On the other hand, $\langle f_{{\rm SNr},i}^{(Ch)} \rangle$ in the channels 2 and 3 becomes increased a little to 27.3 and 28.4 Hz, respectively, because of increased inter-channel IP synaptic current from the channel 1, resulting in suppressing competing actions in the channels 2 and 3. In the last 3rd time interval $I_3$, $\langle f_{{\rm SNr},i}^{(2)} \rangle$ in the 2nd channel (receiving cortical input with $f_{\rm Ctx}^{(2)}=23$ Hz) becomes reduced to 2.4 Hz, leading to action selection in the channel 2. In contrast, $\langle f_{{\rm SNr},i}^{(Ch)} \rangle$ in the channels 1 and 3 becomes increased to 6.9 and 29.8 Hz, respectively, due to increase in the inter-channel IP synaptic current from the channel 2.  Thus, action deselection in the channel 1 occurs, resulting in action switching from the channel 1 to channel 2. Also, no action selection in the channel 3 occurs repeatedly.

\begin{figure*}
\includegraphics[width=1.4\columnwidth]{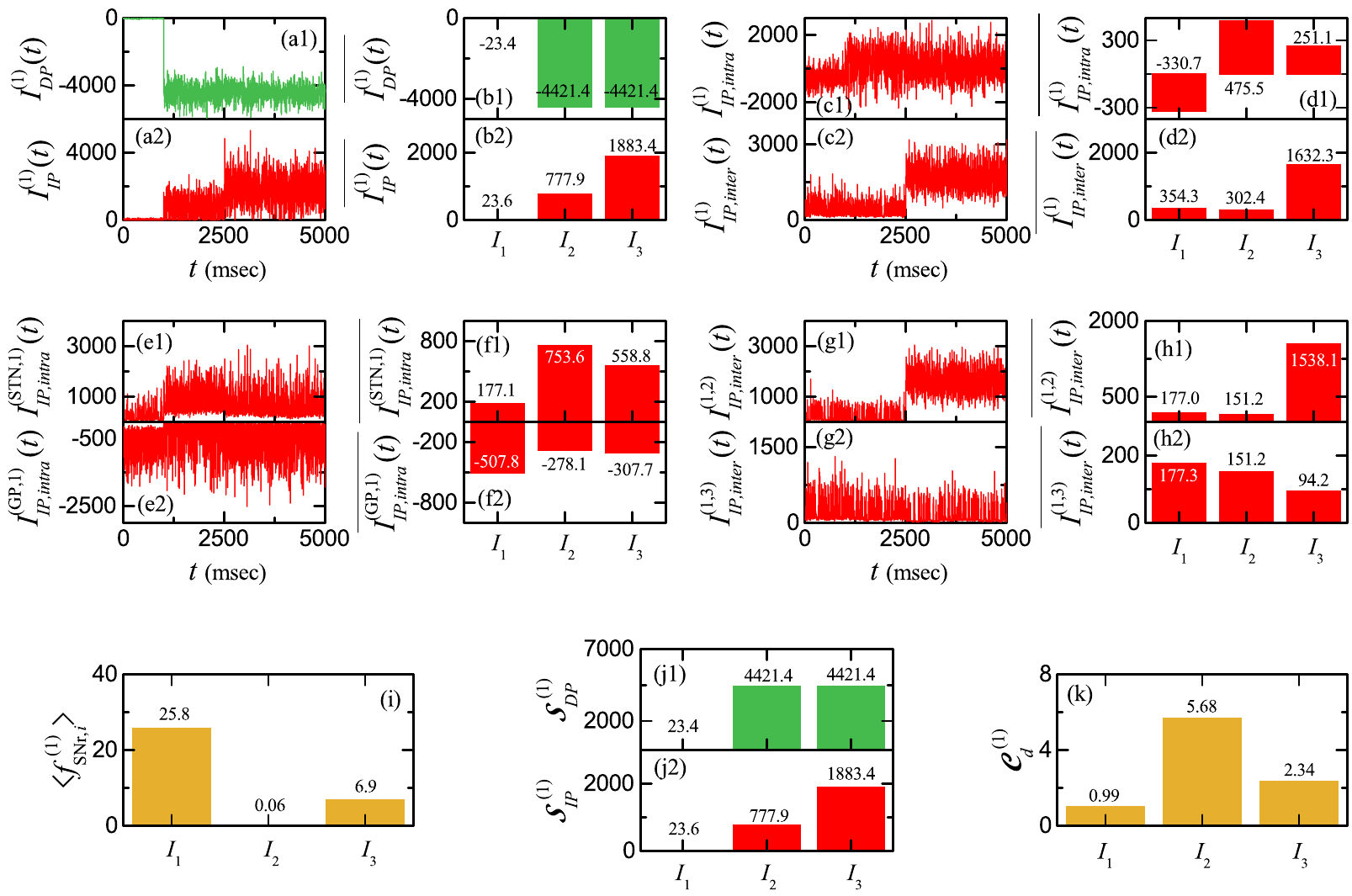}
\caption{Quantitative analysis for action selection in the channel 1 via DP (green) and IP (red) synaptic currents into the channel 1.
Time series of (a1) DP synaptic current $I_{DP}^{(1)}$ (t) and (a2) IP synaptic current $I_{IP}^{(1)} (t)$ versus $t$. Interval-averaged (b1) DP synaptic current $\overline{I_{DP}^{(1)}}$ and (b2) IP synaptic current $\overline{I_{IP}^{(1)}}$ in the time intervals $I_1$, $I_2$, and $I_3$.
(c)-(d) Decomposition of $I_{IP}^{(1)} (t)$ into the intra- and inter-channel IP synaptic currents. Time series of (c1) intra-channel IP synaptic current $I_{IP,intra}^{(1)}(t)$ and (c2) inter-channel IP synaptic current $I_{IP,inter}^{(1)}(t)$ versus $t$. Interval-averaged (d1) intra-channel IP synaptic current $\overline{I_{IP,intra}^{(1)}}$ and (d2) inter-channel IP synaptic current $\overline{I_{IP,inter}^{(1)}}$ in the time intervals $I_1$, $I_2$, and $I_3$.
(e)-(f) Decomposition of $I_{IP,intra}^{(1)} (t)$ into the intra-channel IP synaptic currents from STN and GP. Time series of intra-channel IP synaptic current from (e1) STN $I_{IP,intra}^{({\rm STN,1})}(t)$ and (e2) GP $I_{IP,intra}^{({\rm GP,1})}(t)$ versus $t$. Interval-averaged intra-channel IP synaptic current from (f1) STN $\overline{I_{IP,intra}^{({\rm STN,1})}}$ and (f2) GP $\overline{I_{IP,intra}^{({\rm GP,1})}}$ in the time intervals $I_1$, $I_2$, and $I_3$.
(g)-(h) Decomposition of $I_{IP,inter}^{(1)}(t)$ into the inter-channel IP synaptic currents from the channels 2 and 3 into the channel 1. Time series of inter-channel IP synaptic current from STN in (g1) the channel 2 $I_{IP,inter}^{({\rm 1,2})} (t)$ and (g2) the channel 3 $I_{IP,intra}^{({\rm 1,3})} (t)$ versus $t$. Interval-averaged inter-channel IP synaptic current from STN in (h1) the channel 2 $\overline{I_{IP,inter}^{({\rm 1, 2})}}$ and (h2) the channel 3 $\overline{I_{IP,intra}^{({\rm 1, 3})}}$ in the time intervals $I_1$, $I_2$, and $I_3$.
(i) Histograms of population-averaged mean firing rates $\langle f_{{\rm SNr},i}^{(1)} \rangle$ of SNr neurons in the channel 1 for each time intervals $I_1$, $I_2$, and $I_3$. Histograms of interval-averaged strengths of (j1) DP (${\cal S}_{DP}^{(1)}$) and (j2) IP (${\cal S}_{IP}^{(1)}$) and (k) competition degree
${\cal C}_d^{(1)}$ in each time intervals $I_1$, $I_2$, and $I_3$. Units of currents and MFRs are pA and Hz, respectively.
}
\label{fig:Ch1}
\end{figure*}

\subsection{Action Selection and Switching}
\label{subsec:ASS}
Figure \ref{fig:ASS} shows diagrams for action selection and switching through change in the DP synaptic current $I_{DP}^{(Ch)}(t)$ and the IP synaptic current
$I_{IP}^{(Ch)}(t)$ into the SNr (output nucleus) in a channel in the time intervals, $I_1$, $I_2,$ and $I_3.$
D1 SPNs in the channel provide focused inhibition to the SNr directly via the DP synaptic current, $I_{DP}^{(Ch)}(t)$.
On the other hand, D2 SPNs are linked indirectly to the SNr via IP, crossing the STN and the GP.
Thus, intra-channel IP synaptic currents $I_{IP,intra}^{(Ch)}(t)$ from the source nuclei, STN and GP, are given to the SNr in the same channel;
\begin{equation}
I_{IP,intra}^{(Ch)}(t) = I_{IP,intra}^{({\rm STN},Ch)}(t) + I_{IP,intra}^{({\rm GP},Ch)}(t).
\label{eq:Intra}
\end{equation}
We also note that, inter-channel connections are made through diffusive excitation from the STN in a channel to the target nuclei, SNr and GP, in all the 3 channels \cite{Parent1,Parent2,Parent3}. Thus, inter-channel IP synaptic currents $I_{IP,inter}^{(Ch)}(t)$ from the STN in the two neighboring channels
are provided to the SNr in a channel (Ch);
\begin{equation}
I_{IP,inter}^{(Ch)}(t)= I_{IP,inter}^{(Ch, Ch')}(t) + I_{IP,inter}^{(Ch, Ch'')}(t),
\label{eq:Inter}
\end{equation}
where $Ch'$ and $Ch''$ are neighboring channels. In this way, the (total) IP synaptic current is composed of the intra- and inter-channel IP currents;
\begin{equation}
I_{IP}^{(Ch)}(t) = I_{IP,intra}^{(Ch)}(t) + I_{IP,inter}^{(Ch)}(t).
\label{eq;IP}
\end{equation}

Firing activity (i.e., population-averaged MFRs, $\langle f_i^{({\rm SNr})} \rangle$) of the output nucleus, SNr, is determined via competition between the DP synaptic current $I_{DP}^{(Ch)}(t)$  and the IP synaptic current $I_{IP}^{(Ch)}(t)$. Their competition may be well characterized in terms of recently-introduced competition degree ${\cal C}_d^{(Ch)}$ in our prior work \cite{KimPD}. ${\cal C}_d^{(Ch)}$ is given by the ratio of strength [${\cal S}_{DP}^{(Ch)} (= |I_{DP}^{(Ch)}(t)|)$] of $I_{DP}^{(Ch)}(t)$ to strength [${\cal S}_{IP}^{(Ch)} (=|I_{IP}^{(Ch)}(t)|)$] of $I_{IP}^{(Ch)}(t)$:
\begin{equation}
{\cal C}_d^{(Ch)}  =  \frac { {\cal S}_{DP}^{(Ch)} } { {\cal S}_{IP}^{(Ch)} }.
\label{eq:CD}
\end{equation}
Thus, ${\cal C}_d^{(Ch)}$ plays a good role of indicator for the synaptic inputs into the SNr, in contrast to the output indicator,
$\langle f_{{\rm SNr},i}^{(Ch)} \rangle$. Hence, relationship between ${\cal C}_d^{(Ch)}$ and $\langle f_{{\rm SNr},i}^{(Ch)} \rangle$ may be regarded as the cause-and-effect. The larger ${\cal C}_d^{(Ch)}$ is, the lower $\langle f_{{\rm SNr},i}^{(Ch)} \rangle$ of the SNr neurons becomes.
In the channel with the lowest $\langle f_{{\rm SNr},i}^{(Ch)} \rangle$, the BG gate to the thalamus is open (i.e., the thalamus becomes disinhibited).
Consequently, a desired action may be selected in the channel with the largest ${\cal C}_d^{(Ch)}$ where $\langle f_{{\rm SNr},i}^{(Ch)} \rangle$ of the SNr neurons is the lowest. Hereafter, we employ the competition degree ${\cal C}_d^{(Ch)}$ (input indicator) to determine a desired action selection, instead of the population-averaged MFR of the SNr neurons $\langle f_{{\rm SNr},i}^{(Ch)} \rangle$ (output indicator).

Figure \ref{fig:ASS}(a1) shows the case of the time interval $I_1$; tonic cortical inputs of 3 Hz are provided to all the 3 channels.
The channel 1 is at the center, while the other neighboring channels 2 and 3 are on the right and the left sides, respectively.
We focus on the (center) channel 1. Focused inhibition from D1 SPNs is directly provided to the SNr via the DP synaptic current (green), $I_{DP}^{(1)}$.
On the other hand, D2 SPNs are indirectly linked to the SNr via IP, crossing the intermediate nuclei, STN and GP.
There are two kinds of IP synaptic currents (red), $I_{IP}^{(1)}$; intra-channel IP synaptic current ($I_{IP,intra}^{(1)}$) and
inter-channel IP synaptic currents ($I_{IP,inter}^{(1)}$) from the neighboring channels.
In the case of $I_{IP,intra}^{(1)}$, there are two sources, STN and GP, in the same channel; $I_{IP,intra}^{(1)} = I_{IP,intra}^{({\rm STN}, 1)} + I_{IP,intra}^{({\rm GP}, 1)}.$ Also, $I_{IP,inter}^{(1)}$ consists of two inter-channel IP synaptic currents from the channels 2 and 3;
$I_{IP,inter}^{(1)}= I_{IP,inter}^{(1, 2)} + I_{IP,inter}^{(1, 3)}$. Strengths of all these 5 synaptic currents are denoted by widths of their lines.
In this case of time interval $I_1,$ strength of the DP current (i.e., $|I_{DP}^{(1)}|$) is nearly the same as that of the IP current (i.e., $|I_{IP}^{(1)}|$).
Due to balance between DP and IP in each channel (i.e., ${\cal C}_d^{(Ch)} \simeq 1$), the SNr fires very actively, leading to inhibition of the thalamus. Consequently, no action selection is made in the case of tonic cortical inputs to the 3 channels.

Next, we consider the case of the 2nd time interval $I_2$ in Fig.~\ref{fig:ASS}(a2); cortical input with the frequency $f_{\rm Ctx}^{(1)}=15$ Hz is provided to the channel 1, while the other channels 2 and 3 receive tonic cortical inputs of 3 Hz. The total IP synaptic current into the channel 1, $I_{IP}^{(1)}$, is given by the sum of the intra- and inter-channel IP synaptic currents (i.e., $I_{IP}^{(1)} = I_{IP,intra}^{(1)} + I_{IP,inter}^{(1)}$). Due to the salient cortical input to the channel 1, strength of the DP current, $|I_{DP}^{(1)}|$, becomes larger than that of the IP current, $|I_{IP}^{(1)}|$.
Due to strong focused inhibition via $I_{DP}^{(1)},$  the competition degree ${\cal C}_d^{(1)}$ of the channel 1 increases much more than that
(${\cal C}_d^{(1)} \simeq 1$) in the case of $I_1$ where the channel 1 receives the tonic cortical input. Consequently, firing activity of the SNr becomes so much reduced (off-center effect). Thus, the thalamus becomes disinhibited, leading to action selection in the channel 1. The action-selected channel 1 is shaded in yellow color.

But, in the 3rd time interval $I_3$, action switching from the channel 1 to the channel 2 occurs, as shown in in Fig.~\ref{fig:ASS}(a3).
In this case, the channel 2 receives more salient cortical input with frequency $f_{\rm Ctx}^{(2)} = 23$ Hz than that ($f_{\rm Ctx}^{(1)} = 15$ Hz) in the case of the channel 1; the channel 3 continues to receive the tonic cortical input of 3 Hz. In this case, due to the strong DP synaptic current
$I_{DP}^{(2)},$ the competition degree ${\cal C}_d^{(2)}$ of the channel 2 becomes increased much more than that (${\cal C}_d^{(2)} \simeq 1$) in the case of $I_2$ where the channel 2 receives the tonic cortical input. On the other hand, the competition degree ${\cal C}_d^{(1)}$ of the channel 1 becomes decreased much in comparison to that in $I_2$, because of the strong inter-channel IP synaptic current, $I_{IP,inter}^{(1, 2)}(t)$, from the channel 2 to the channel 1 [see Fig.~\ref{fig:ASS}(b)].
Thus, the competition degree ${\cal C}_d^{(2)}$ of the channel 2 becomes larger than ${\cal C}_d^{(1)}$ of the channel 1, resulting in action selection in the channel 2 (shaded in yellow).

Figure \ref{fig:ASS}(b) shows action deselection in the channel 1 in the time interval $I_3$. We note that, due to strong inter-channel IP synaptic current
$I_{IP,inter}^{(1, 2)}(t)$ from the channel 2, the competition degree ${\cal C}_d^{(1)}$ of the channel 1 becomes reduced, leading to suppress the competing action
in the channel 1 (on-surround effect causing contrast enhancement). As a result, action deselection occurs in the channel 1, and the action selection in the channel 2 becomes highlighted due to contrast enhancement.

\subsection{Quantitative Analysis for Action Selection in The Channel 1}
\label{sunsec:QACh1}
From now on, we make quantitative analysis of functions of the DP and the two intra- and inter-channel IPs for action selection explicitly.
Figure \ref{fig:Ch1} shows quantitative analysis for action selection in the channel 1 via DP (green) and IP (red) synaptic currents into the channel 1.
Detailed data, associated with the DP and the IP synaptic currents, are given in Figs.~\ref{fig:Ch1}(a1)-\ref{fig:Ch1}(h2).
(Units of DP and IP synaptic currents are pA; for simplicity, we omit the unit when presenting values of currents.)

Time series of the DP synaptic current $I_{DP}^{(1)}(t)$ [see Fig.~\ref{fig:Ch1}(a1); green color] and the IP synaptic current $I_{IP}^{(1)}(t)$ [see Fig.~\ref{fig:Ch1}(a2); red color] are shown in the time interval of $0 < t < 5,000$ msec; $I_{DP}^{(1)}(t)$ (green) and $I_{IP}^{(1)}(t)$ (red) are also shown
in Figs.~\ref{fig:ASS}(a1) and \ref{fig:ASS}(a2). Their interval-averaged DP and IP synaptic currents, $\overline{I_{DP}^{(1)}(t)}$ and $\overline{I_{IP}^{(1)}(t)}$, in each time interval, $I_1,$ $I_2,$ and $I_3,$ are also given in Figs.~\ref{fig:Ch1}(b1) and \ref{fig:Ch1}(b2), respectively; the overline denotes time averaging.

More details on $I_{IP}^{(1)}(t)$ are also provided. We first decompose $I_{IP}^{(1)}(t)$ into its components, the intra- and inter-channel IP synaptic currents, $I_{IP,intra}^{(1)}(t)$ and $I_{IP,inter}^{(1)}(t)$. Their time series and interval averaged ones are shown in Figs.~\ref{fig:Ch1}(c1)-\ref{fig:Ch1}(c2) and Figs.~\ref{fig:Ch1}(d1)-\ref{fig:Ch1}(d2), respectively. We also make one more decomposition for $I_{IP,intra}^{(1)}(t)$ and $I_{IP,inter}^{(1)}(t)$. In the case of $I_{IP,intra}^{(1)}(t)$, there are two sources in the same channel 1, STN and GP. Figures \ref{fig:Ch1}(e1)-\ref{fig:Ch1}(e2) and Figures \ref{fig:Ch1}(f1)-\ref{fig:Ch1}(f2) show the time series of $I_{IP,intra}^{({\rm STN}, 1)}(t)$ and $I_{IP,intra}^{({\rm GP}, 1)}(t)$ [see Figs.~\ref{fig:ASS}(a1) and \ref{fig:ASS}(a2); red color] and their interval-averaged ones, $\overline{I_{IP,intra}^{({\rm STN}, 1)}(t)}$ and $\overline{I_{IP,intra}^{({\rm GP}, 1)}(t)},$ respectively. For the case of $I_{IP,inter}^{(1)}(t)$, there are two inter-channel IP synaptic currents,
$I_{IP,inter}^{(1, 2)}(t)$ and $I_{IP,inter}^{(1, 3)}(t)$ from the neighboring channels 2 and 3 [see Figs.~\ref{fig:ASS}(a1) and \ref{fig:ASS}(a2); red color].
Their time series and interval-averaged ones are given in Figs.~\ref{fig:Ch1}(g1)-\ref{fig:Ch1}(g2) and Figs.~\ref{fig:Ch1}(h1)-\ref{fig:Ch1}(h2), respectively.

At $t=1,000$ msec, cortical input with $f_{\rm Ctx}^{(1)} = 15$ Hz starts to be provided to the channel 1 [see Fig.~\ref{fig:PIFB}(a)].
Thus, in the time intervals $I_2$ and $I_3$, strong focused inhibition from D1 SPNs is given to the SNr via the DP synaptic current $I_{DP}^{(1)}(t)$ (green), as shown in Figs.~\ref{fig:Ch1}(a1) and \ref{fig:Ch1}(b1). We note that, the interval-averaged DP synaptic current $\overline{I_{DP}^{(1)}(t)}$ in $I_2$ and $I_3$ is -4,421.4, in contrast to that (= -23.4) in $I_1$ for the tonic cortical input of 3 Hz. Thus, the (inhibitory) DP synaptic current $I_{DP}^{(1)}(t)$ suppresses strongly the firing activity of the SNr in $I_2$ [see the population-averaged MFR of the SNr neurons $\langle f_{{\rm SNr},i}^{(1)} \rangle$ (= 0.06 Hz) in $I_2$
in Fig.~\ref{fig:PIFB}(b)], leading to disinhibition of the thalamus, which is in contrast to the case of $I_1$ (tonic cortical input of 3 Hz) with
$\langle f_{{\rm SNr},i}^{(1)} \rangle$ = 25.8 Hz, resulting in inhibition of the SNr.

In addition to $I_{DP}^{(1)}(t)$, the IP synaptic current $I_{IP}^{(1)}(t)$ (red) is also provided to the SNr. Its time series and interval-averaged one ($\overline{ I_{IP}^{(1)}(t)}$) are given in Figs.~\ref{fig:Ch1}(a2) and \ref{fig:Ch1}(b2), respectively. Its interval-averaged value jumps from 23.6 in $I_1$ to 777.9 in $I_2$ due to cortical input of 15 Hz in the channel 1. Since $I_{IP}^{(1)}(t)$  is an excitatory current, it enhances the firing activity of the SNr, in contrast to the case of $I_{DP}^{(1)}(t)$, resulting in inhibition of the thalamus. But, we note that $I_{IP}^{(1)}(t)$ is much less than the magnitude of $I_{DP}^{(1)}(t).$ Hence, the net firing activity of the SNr neurons becomes so much reduced to 0.06 Hz in $I_2,$ as shown in Fig.~\ref{fig:PIFB}(b).

We also decompose $I_{IP}^{(1)}(t)$ into its components. In the case of $I_2$, the (excitatory) interval-averaged intra-channel IP synaptic current from STN (receiving the cortical input of 15 Hz), $\overline{ I_{IP,intra}^{({\rm STN}, 1)}(t) },$ increases to 753.6 [see Fig.~\ref{fig:Ch1}(f1)]. On the other hand, the
magnitude of (inhibitory) interval-averaged intra-channel IP synaptic current from GP, $| \overline{ I_{IP,intra}^{({\rm GP}, 1)}(t) }|$, decreases to 278.1 in $I_2$ [see Fig.~\ref{fig:Ch1}(f2)] because D2 SPN (receiving the cortical input of 15 Hz) inhibits the GP more in $I_2$ than in $I_1.$ Thus, the (total) interval-averaged intra-channel IP synaptic current $\overline{ I_{IP,intra}^{(1)}(t) }$ becomes increased to 475.5 in $I_2$ [see Fig.~\ref{fig:Ch1}(d1)]. $I_{IP,intra}^{(1)}(t)$ is a major contribution to increase in $I_{IP}^{(1)}(t)$ in $I_2.$

In contrast to $I_{IP,intra}^{(1)}(t)$, the interval-averaged inter-channel IP synaptic current $I_{IP,inter}^{(1)}(t)$ in $I_2$ is found to decrease to 302.4 than that (= 354.3) in $I_1$ [see Fig.~\ref{fig:Ch1}(d2)], because the interval-averaged inter-channel IP synaptic current from the neighboring channels 2 and 3,
$I_{IP,inter}^{(1, 2)}(t)$ and $I_{IP,inter}^{(1, 3)}(t),$ are decreased to 151.2 in $I_2$ [see Figs.~\ref{fig:Ch1}(h1) and \ref{fig:Ch1}(h2)].
In $I_2,$ the STN in the channel 1 (receiving cortical input of 15 Hz) makes diffusive projections to the GP and the SNr in the channels 2 and 3. Thus, the firing activity of the GP in the channels 2 and 3 becomes increased, leading to decrease in the firing activity of the STN in the channels 2 and 3. Consequently,
both $I_{IP,inter}^{(1, 2)}(t)$ and $I_{IP,inter}^{(1, 3)}(t)$ becomes decreased, resulting in decrease in
$\overline{ I_{IP,inter}^{(1)}(t) }$ to 302.4 in $I_2$, and $I_{IP,inter}^{(1)}(t)$ becomes a minor contribution to increase in $I_{IP}^{(1)}(t)$ in $I_2.$

\begin{figure*}
\includegraphics[width=1.4\columnwidth]{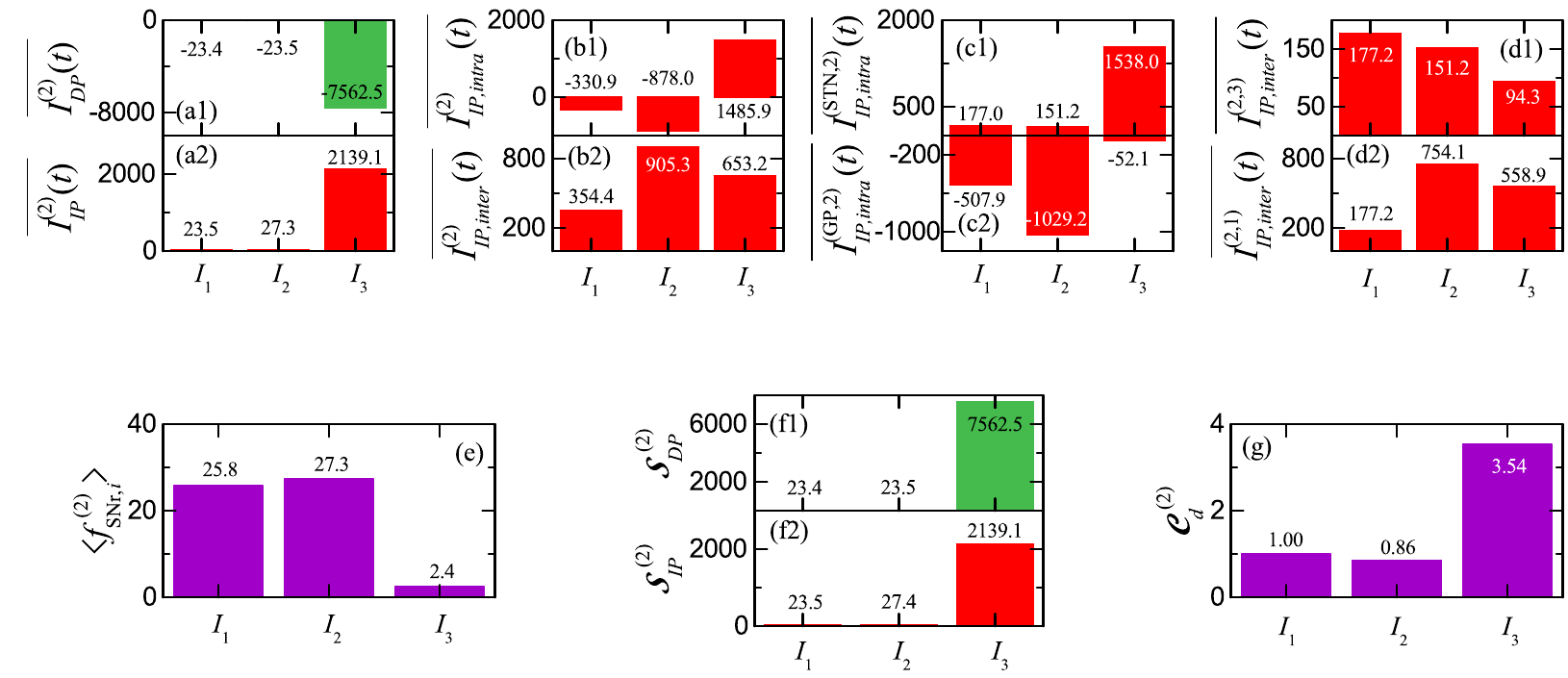}
\caption{Quantitative analysis for action selection in the channel 2 through DP (green) and IP (red) synaptic currents into the channel 2.
Interval-averaged (a1) DP synaptic current $\overline{I_{DP}^{(2)}}$ and (a2) IP synaptic current $\overline{I_{IP}^{(2)}}$ in the time intervals $I_1$, $I_2$, and $I_3$. (b) Decomposition of $I_{IP}^{(2)} (t)$ into the intra- and inter-channel IP synaptic currents. Interval-averaged (b1) intra-channel IP synaptic current $\overline{I_{IP,intra}^{(2)}}$ and (b2) inter-channel IP synaptic current $\overline{I_{IP,inter}^{(2)}}$ in the time intervals $I_1$, $I_2$, and $I_3$.
(c) Decomposition of $I_{IP,intra}^{(2)} (t)$ into the intra-channel IP synaptic currents from STN and GP. Interval-averaged intra-channel IP synaptic current from (c1) STN $\overline{I_{IP,intra}^{({\rm STN, 2})}}$ and (c2) GP $\overline{I_{IP,intra}^{({\rm GP,2})}}$ in the time intervals $I_1$, $I_2$, and $I_3$.
(d) Decomposition of $I_{IP,inter}^{(2)}(t)$ into the inter-channel IP synaptic currents from the channel 3 and 1 into the channel 2. Interval-averaged inter-channel IP synaptic current from STN in (d1) the channel 3 $\overline{I_{IP,inter}^{({\rm 2,3})}}$ and (d2) the channel 1 $\overline{I_{IP,intra}^{({\rm 2,1})}}$ in the time intervals $I_1$, $I_2$, and $I_3$.
(e) Histograms of population-averaged mean firing rates $\langle f_{{\rm SNr},i}^{(2)} \rangle$ of SNr neurons in the channel 2 for each time intervals $I_1$, $I_2$, and $I_3$. Histograms of interval-averaged strengths of (f1) DP (${\cal S}_{DP}^{(2)}$) and (f2) IP (${\cal S}_{IP}^{(2)}$) and (g) competition degree
${\cal C}_d^{(2)}$ in each time intervals $I_1$, $I_2$, and $I_3$. Units of currents and MFRs are pA and Hz, respectively.
}
\label{fig:Ch2}
\end{figure*}

We now consider the next time interval $I_3$ ($2,500 < t < 5,000$ msec). The interval-averaged DP synaptic current $\overline {I_{DP}^{(1)}(t)}$ in $I_3$ is the same as that (= -4,421.4) in $I_2$, because the same cortical input of 15 Hz is given to the channel 1 in $I_3.$ But, the interval-averaged IP synaptic current $\overline {I_{IP}^{(1)}(t)}$ is found to increase to 1,883.4 more than that (= 777.9) in $I_2$ [see Fig.~\ref{fig:Ch1}(b2)], mainly due to increase in the inter-channel IP synaptic current from the channel 2, $I_{IP,inter}^{(1, 2)}(t)$. We note that, at $t = 2,500$ msec, cortical input with frequency $f_{\rm Ctx}^{(2)} = 23$ Hz begins to be provided to the channel 2. Then, the STN (receiving this cortical input of 23 Hz) in the channel 2 makes diffusive excitatory projection to the SNr and the GP in the neighboring channels 1 and 3. Thus, the interval-averaged inter-channel IP synaptic current $\overline {I_{IP,inter}^{(1)}(t)}$ in $I_3$ becomes increased to 1632.3 in $I_3,$ due to increase in the inter-channel IP synaptic current from the channel 2, $I_{IP,inter}^{(1, 2)}(t)$ [see Figs.~\ref{fig:Ch1} (h1) and \ref{fig:Ch1} (d2)]. Because of increase in $I_{IP}^{(1)}(t)$ in $I_3$, the net firing activity of the SNr
($\langle f_{{\rm SNr},i}^{(1)} \rangle = $ 6.9 Hz) in $I_3$ becomes more enhanced than that ($\langle f_{{\rm SNr},i}^{(1)} \rangle=$ 0.06 Hz) in $I_2$ [see Fig.~\ref{fig:PIFB}(b)].

Firing activity of the SNr (output nucleus) is determined through competition between the above DP and IP synaptic currents into the SNr, and
it may be well characterized in terms of their population-averaged MFRs $\langle f_{{\rm SNr},i}^{(1)} \rangle$ of Eq.~(\ref{eq:PMFR}). Thus, $\langle f_{{\rm SNr},i}^{(1)} \rangle$ becomes a good indicator of the output activity of the BG. Figure \ref{fig:Ch1}(i) shows population-averaged MFRs $\langle f_{{\rm SNr},i}^{(1)} \rangle$ of the SNr neurons in the channel 1 for the time intervals $I_1,$ $I_2,$ and $I_3.$ In $I_1$ (where the channel 1 receives the tonic cortical input of 3 Hz), strengths of the DP and IP synaptic currents are nearly the same, and the SNr neurons fire very actively with $\langle f_{{\rm SNr},i}^{(1)} \rangle=$ 25.8 Hz. Then, the BG gate to the thalamus becomes locked, leading to inhibition of the thalamus. But, in $I_2$ (where the channel 1 receives the cortical input of 15 Hz), the (inhibitory) DP synaptic current is stronger than the (excitatory) IP synaptic current, and hence $\langle f_{{\rm SNr},i}^{(1)} \rangle$ becomes so much reduced to 0.06 Hz. In this case, the BG gate to the thalamus becomes opened, resulting in disinhibition of the thalamus. In $I_3,$ strength of the IP synaptic current becomes larger than that in $I_2,$ and hence $\langle f_{{\rm SNr},i}^{(1)} \rangle$  becomes increased to 6.9 Hz.

The above population-averaged MFR $\langle f_{{\rm SNr},i}^{(1)} \rangle$ of the SNr neurons is determined via competition between the DP synaptic current
$I_{DP}^{(1)}(t)$ and the IP synaptic current $I_{IP}^{(1)}(t)$ into the SNr in the channel 1. Their competition may be well characterized in terms of their competition degree ${\cal C}_d^{(1)}$ of Eq.~(\ref{eq:CD}), given by the ratio of strength ${\cal S}_{DP}^{(1)}$ (= $|I_{DP}^{(1)}(t)|$) of the DP synaptic current to strength ${\cal S}_{IP}^{(1)}$ (= $|I_{IP}^{(1)}(t)|$) of the IP synaptic current. Thus, ${\cal C}_d^{(1)}$ becomes a good indicator for synaptic inputs
into the SNr, in contrast to the output indicator, $\langle f_{{\rm SNr},i}^{(1)} \rangle$. Consequently, relationship between ${\cal C}_d^{(1)}$ and $\langle f_{{\rm SNr},i}^{(1)} \rangle$ could be regarded as the cause-and-effect. The larger ${\cal C}_d^{(1)}$ is, the lower $\langle f_{{\rm SNr},i}^{(1)} \rangle$ becomes.

Figures \ref{fig:Ch1}(j1) and \ref{fig:Ch1}(j2) show ${\cal S}_{DP}^{(1)}$ and ${\cal S}_{IP}^{(1)}$ in each time interval, $I_1,$ $I_2,$ and $I_3$, respectively. In $I_1$ (tonic cortical input of 3 Hz), ${\cal S}_{DP}^{(1)}$ and ${\cal S}_{IP}^{(1)}$ are nearly the same. But, in $I_2$ (cortical input of 15 Hz) ${\cal S}_{DP}^{(1)}$ is much larger than ${\cal S}_{IP}^{(1)}$, mainly due to focused inhibition from D1 SPNs to the SNr via DP. In $I_3$ (where the channel 2 receives cortical input of 23 Hz), ${\cal S}_{IP}^{(1)}$ increases because of increased inter-channel IP synaptic current $I_{IP,inter}^{(1,2)}$ from the channel 2.
Then, the competition degree ${\cal C}_d^{(1)}$ of the channel 1 is given in Fig.~\ref{fig:Ch1}(k).
In $I_1$ (where the channel 1 receives tonic cortical input of 3 Hz), ${\cal C}_d^{(1)} = 0.99$ (i.e., DP and IP are nearly balanced), leading to no action selection (i.e., BG gate to the thalamus is locked); refer to Fig.~\ref{fig:ASS}(a1). But, in $I_2$ (where the channel 1 receives cortical input of 15 Hz),
${\cal C}_d^{(1)} = 5.68$ (i.e., DP is 5.68 times stronger than IP), and hence, as a result of the focused inhibition from D1 SPNs to the SNr via DP (off-center effect), the BG gate to the thalamus becomes open, leading to action selection; refer to Fig.~\ref{fig:ASS}(a2). In $I_3$ (where the channel 2 receives cortical input of 23 Hz), ${\cal C}_d^{(1)}$ becomes decreased to 2.34 due to the increased inter-channel IP current $I_{IP,inter}^{(1, 2)}$ from the channel 2
[see Fig.~\ref{fig:Ch1}(h1)]. In this case, action deselection of the channel 1 occurs, as shown in Fig.~\ref{fig:ASS}(b), because the increased inter-channel IP current $I_{IP,inter}^{(1, 2)}$ from the channel 2 suppresses the competing action in the channel 1 (on-surround effect, causing contrast enhancement to highlight action selection in the channel 2). Consequently, action switching takes place in $I_3$ from the channel 1 to the channel 2 with larger ${\cal C}_d^{(2)}$ (= 3.54).

\subsection{Quantitative Analysis for Action Selection in The Channel 2}
\label{subsec:QACh2}
We now make quantitative analysis for action selection in the channel 2 which receives the cortical input of 23 Hz in $I_3.$
Figure \ref{fig:Ch2} shows quantitative analysis for action selection in the channel 2 via DP (green) and IP (red) synaptic currents into the channel 2.
Detailed data, related to the DP [see Fig.~\ref{fig:ASS}(a3); green] and the IP [see Fig.~\ref{fig:ASS}(a3); red] synaptic currents, $I_{DP}^{(2)}(t)$ and $I_{IP}^{(2)}(t)$, are given in Figs.~\ref{fig:Ch2}(a1)-\ref{fig:Ch2}(d2). Interval-averaged DP and IP synaptic currents, $\overline{I_{DP}^{(2)}(t)}$ and $\overline{I_{IP}^{(2)}(t)}$, in each time interval, $I_1,$ $I_2,$ and $I_3,$ are shown in Figs.~\ref{fig:Ch2}(a1) and \ref{fig:Ch2}(a2), respectively.

We also decompose $I_{IP}^{(2)}(t)$ into its components, the intra- and inter-channel IP synaptic currents, $I_{IP,intra}^{(2)}(t)$ and $I_{IP,inter}^{(2)}(t)$. Their interval-averaged ones, $\overline{ I_{IP,intra}^{(2)}(t) }$ and $\overline{ I_{IP,inter}^{(2)}(t)},$ are shown in Figs.~\ref{fig:Ch2}(b1)-\ref{fig:Ch2}(b2). One more decompositions of $I_{IP,intra}^{(2)}(t)$ and $I_{IP,inter}^{(2)}(t)$ are made. For $I_{IP,intra}^{(2)}(t)$, there are two intra-channel IP synaptic currents from the STN and the GP in the same channel 2, $I_{IP,intra}^{({\rm STN}, 2)}(t)$ and $I_{IP,intra}^{({\rm GP}, 2)}(t)$ [see Fig.~\ref{fig:ASS}(a3); red]. Figures \ref{fig:Ch2}(c1)-\ref{fig:Ch2}(c2) show their interval-averaged ones, $\overline{I_{IP,intra}^{({\rm STN}, 2)}(t)}$ and $\overline{I_{IP,intra}^{({\rm GP}, 2)}(t)},$ respectively. In the case of $I_{IP,inter}^{(2)}(t)$, there are two inter-channel IP synaptic currents,
$I_{IP,inter}^{(2, 3)}(t)$ and $I_{IP,inter}^{(2, 1)}(t)$ from the neighboring channels 3 and 1 [see Fig.~\ref{fig:ASS}(a3); red].
Their interval-averaged ones, $\overline{ I_{IP,inter}^{(2, 3)}(t) }$ and $\overline{ I_{IP,inter}^{(2, 1)}(t) }$
are given in Figs.~\ref{fig:Ch2}(d1)-\ref{fig:Ch2}(d2), respectively.

\begin{figure*}
\includegraphics[width=1.4\columnwidth]{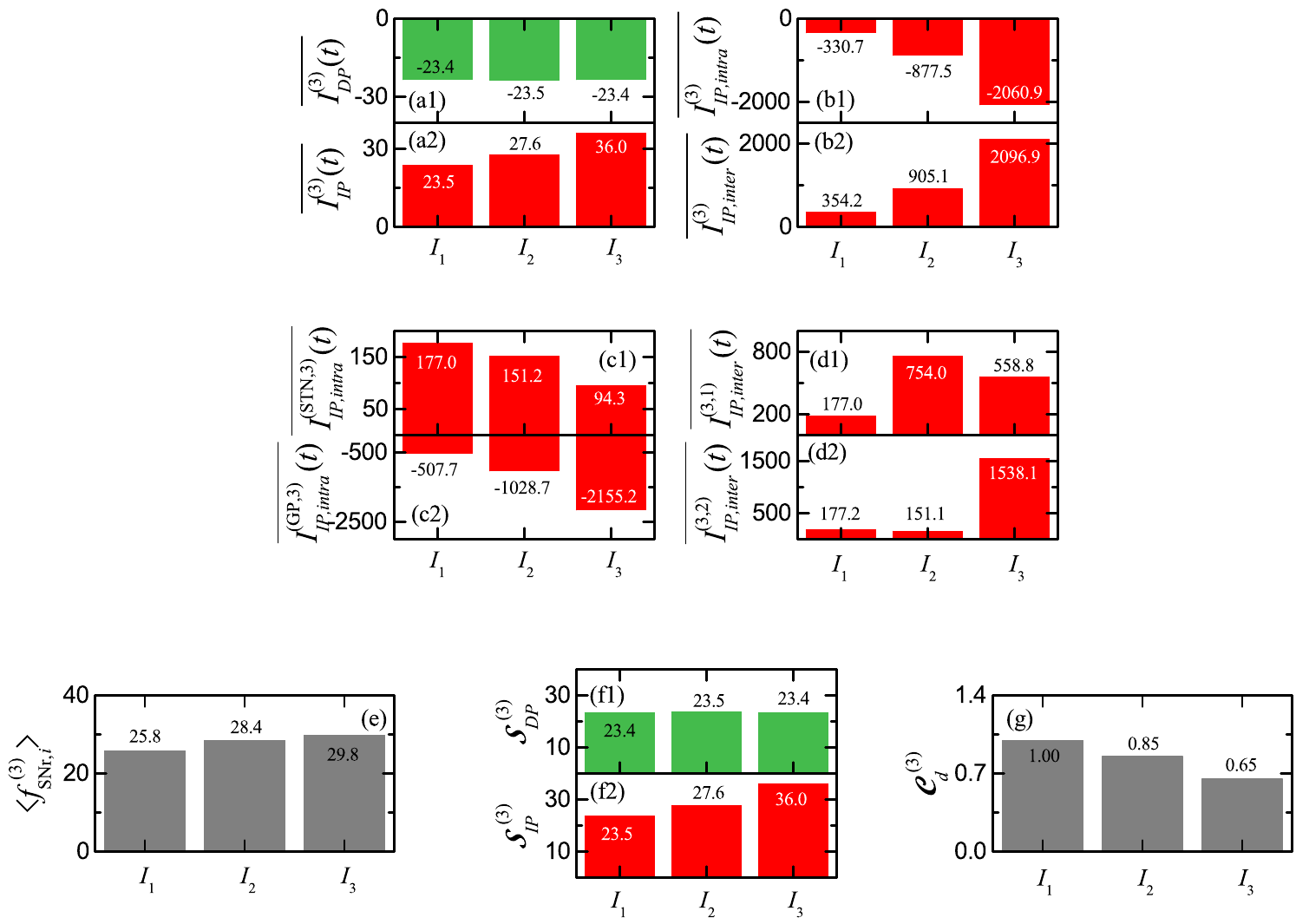}
\caption{Quantitative analysis for action selection in the channel 3 through DP (green) and IP (red) synaptic currents into the channel 3.
Interval-averaged (a1) DP synaptic current $\overline{I_{DP}^{(3)}}$ and (a2) IP synaptic current $\overline{I_{IP}^{(3)}}$ in the time intervals $I_1$, $I_2$, and $I_3$. (b) Decomposition of $I_{IP}^{(3)} (t)$ into the intra- and inter-channel IP synaptic currents. Interval-averaged (b1) intra-channel IP synaptic current $\overline{I_{IP,intra}^{(3)}}$ and (b2) inter-channel IP synaptic current $\overline{I_{IP,inter}^{(3)}}$ in the time intervals $I_1$, $I_2$, and $I_3$.
(c) Decomposition of $I_{IP,intra}^{(3)} (t)$ into the intra-channel IP synaptic currents from STN and GP. Interval-averaged intra-channel IP synaptic current from (c1) STN $\overline{I_{IP,intra}^{({\rm STN, 3})}}$ and (c2) GP $\overline{I_{IP,intra}^{({\rm GP,3})}}$ in the time intervals $I_1$, $I_2$, and $I_3$.
(d) Decomposition of $I_{IP,inter}^{(3)}(t)$ into the inter-channel IP synaptic currents from the channel 1 and 2 into the channel 3. Interval-averaged inter-channel IP synaptic current from STN in (d1) the channel 1 $\overline{I_{IP,inter}^{({\rm 3,1})}}$ and (d2) the channel 2 $\overline{I_{IP,intra}^{({\rm 3,2})}}$ in the time intervals $I_1$, $I_2$, and $I_3$. (e) Histograms of population-averaged mean firing rates $\langle f_{{\rm SNr},i}^{(3)} \rangle$ of SNr neurons in the channel 3 for each time intervals $I_1$, $I_2$, and $I_3$. Histograms of interval-averaged strengths of (f1) DP (${\cal S}_{DP}^{(3)}$) and (f2) IP (${\cal S}_{IP}^{(3)}$) and (g) competition degree ${\cal C}_d^{(3)}$ in each time intervals $I_1$, $I_2$, and $I_3$. Units of currents and MFRs are pA and Hz, respectively.
}
\label{fig:Ch3}
\end{figure*}

At $t=2,500$ msec, cortical input with $f_{\rm Ctx}^{(2)} = 23$ Hz begins to be given to the channel 2 [see Fig.~\ref{fig:PIFB}(a)].
Thus, in the time interval $I_3$, strong focused inhibition from D1 SPNs is provided to the SNr via the DP synaptic current $I_{DP}^{(2)}(t)$ (green), as shown in Fig.~\ref{fig:Ch2}(a1). We note that, the interval-averaged DP synaptic current $\overline{I_{DP}^{(2)}(t)}$ in $I_3$ are -7,562.5, in contrast
to those [= -23.4 ($I_1$) and -23.5 ($I_2$)] in the case of tonic cortical inputs of 3 Hz. Thus, the (inhibitory) DP synaptic current $I_{DP}^{(2)}(t)$ suppresses strongly the firing activity of the SNr in $I_3$ [see the population-averaged MFR of the SNr neurons $\langle f_{{\rm SNr},i}^{(2)} \rangle$ (= 2.4 Hz) in $I_3$ in Fig.~\ref{fig:PIFB}(b)], resulting in disinhibition of the thalamus, which is in contrast to the cases of $I_1$ and $I_2$ (tonic cortical inputs of 3 Hz) with
$\langle f_{{\rm SNr},i}^{(1)} \rangle$ = 25.8 Hz and 27.3 Hz, leading to inhibition of the SNr.

Along with $I_{DP}^{(2)}(t)$, the IP synaptic current $I_{IP}^{(2)}(t)$ (red) is also provided to the SNr.
Its interval-averaged value jumps to 2139.1 in $I_3$ mainly due to strong cortical input of 23 Hz in the channel 2.
The major contribution to $I_{IP}^{(2)}(t)$ in $I_3$ is the intra-channel IP synaptic current [$\overline{I_{IP,intra}^{(2)}(t)}=$ 1,485.9
in Fig.~\ref{fig:Ch2}(b1)]; contribution from the inter-channel IP synaptic current [$\overline{I_{IP,inter}^{(2)}(t)}=$ 653.2
in Fig.~\ref{fig:Ch2}(b2)] is smaller than that of $\overline{I_{IP,intra}^{(2)}(t)}$.
Due to the cortical input of 23 Hz, the intra-channel IP synaptic current from the STN becomes dominant
[$\overline{I_{IP,intra}^{({\rm STN}, 2)}(t)}=$ 1,538.0 in Fig.~\ref{fig:Ch2}(c1)].
For the inter-channel IP synaptic current, the inter-channel IP synaptic current from the channel 1 (which receives 15 Hz cortical input)
is dominant [$\overline{I_{IP,inter}^{(2, 1)}(t)}=$ 558.9 in Fig.~\ref{fig:Ch2}(d2)].
Because $I_{IP}^{(2)}(t)$  is an excitatory current, it enhances the firing activity of the SNr, leading to inhibition of the thalamus,
in contrast to the case of $I_{DP}^{(2)}(t)$. But, we note that $I_{IP}^{(2)}(t)$ is much less than the magnitude of $I_{DP}^{(2)}(t).$ Accordingly, the net firing activity of the SNr neurons becomes much reduced to 2.4 Hz in $I_3,$ as shown in Fig.~\ref{fig:PIFB}(b).

\begin{figure*}
\includegraphics[width=1.2\columnwidth]{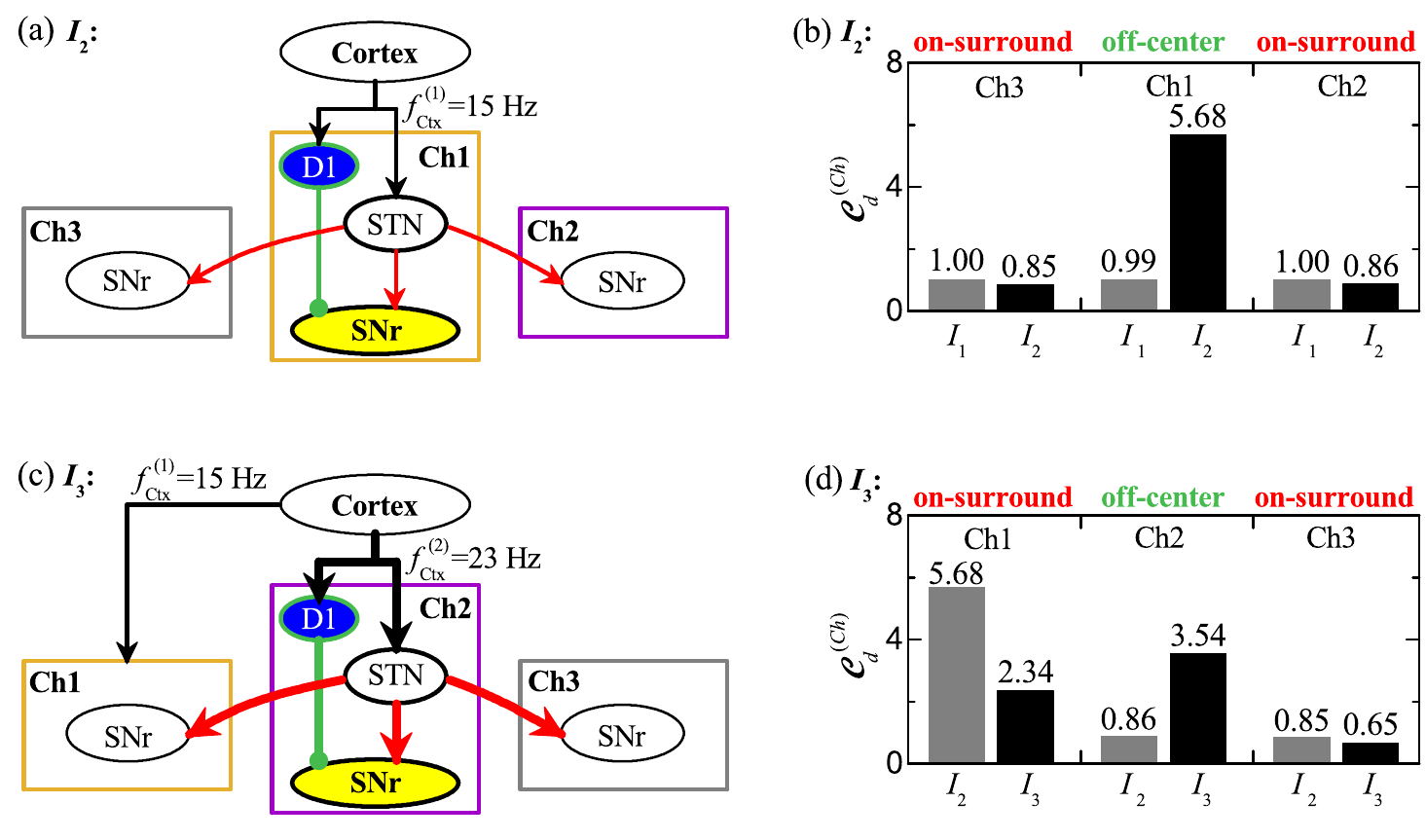}
\caption{Off-center and on-surround effect via DP (green) and IP (red) synaptic currents.
Off-center and on-surround effect in $I_2$. (a) Box diagram for the off-center effect in the channel 1 (receiving cortical input of 15 Hz) via the DP synaptic current (green) from the D1 SPNs to the SNr and the on-surround effect in the neighboring channels 2 and 3 via the inter-channel IP synaptic current from the STN neurons in the channel 1 with selected action in the time interval $I_2$.
(b) Competition degree ${\cal C}_d^{(Ch)}$ for the previous time interval $I_1$ (gray) and the current time interval $I_2$ (black) of the selected channel 1 and the neighboring channels 2 and 3. Increase in ${\cal C}_d^{(1)}$ from 0.99 to 5.68 (off-center effect). Decrease in ${\cal C}_d^{(2)}$ and ${\cal C}_d^{(3)}$
from 1.0 to 0.85 (channel 1) and 0.86 (channel 3).
Off-center and on-surround effect in $I_3$.
(c) Box diagram for the off-center effect in the channel 2 (receiving cortical input of 23 Hz) via the DP synaptic current (green) from the D1 SPNs to the SNr and the on-surround effect in the neighboring channels 3 and 1 via the inter-channel IP synaptic current from the STN neurons in the channel 2 with selected action in  $I_2$. (d) Competition degree ${\cal C}_d^{(Ch)}$ for the previous time interval $I_2$ (gray) and the current time interval $I_3$ (black) of the selected channel 2 and the neighboring channels 3 and 1. Increase in ${\cal C}_d^{(2)}$ from 0.86 to 3.54 (off-center effect). Decrease in ${\cal C}_d^{(3)}$ [${\cal C}_d^{(1)}$]
from 5.68 (0.85) to 2.34 (0.65).
}
\label{fig:OCOS}
\end{figure*}

Figure \ref{fig:Ch2}(e) shows population-averaged MFRs $\langle f_{{\rm SNr},i}^{(2)} \rangle$ of the SNr neurons in the channel 2 for the time intervals $I_1,$ $I_2,$ and $I_3.$ In $I_1$ with tonic cortical input of 3 Hz, strengths of the DP and the IP synaptic currents are nearly the same, and hence the SNr neurons fire very actively with $\langle f_{{\rm SNr},i}^{(2)} \rangle=$ 25.8 Hz, resulting in inhibition of the thalamus. In the next $I_2,$ (where the channel 2 receives the
same tonic cortical input of 3 Hz),  $\langle f_{{\rm SNr},i}^{(2)} \rangle$ increases a little to 27.3 Hz, due to the increased inter-channel IP synaptic current from the channel 1 (with the cortical input of 15 Hz), $I_{IP,inter}^{(2, 1)}(t).$ But, in $I_3$ (where the channel 2 receives the cortical input of 23 Hz),
the (inhibitory) DP synaptic current is stronger than the (excitatory) IP synaptic current, and hence $\langle f_{{\rm SNr},i}^{(2)} \rangle$ becomes much reduced to 2.4 Hz, leading to disinhibition of the thalamus.

Figures \ref{fig:Ch2}(f1) and \ref{fig:Ch2}(f2) show ${\cal S}_{DP}^{(2)}$ (strength of the DP synaptic current) and ${\cal S}_{IP}^{(2)}$ (strength of the IP synaptic current) for the channel 2 in each time interval, $I_1,$ $I_2,$ and $I_3$, respectively. In $I_1$ (tonic cortical input of 3 Hz), ${\cal S}_{DP}^{(2)}$ and ${\cal S}_{IP}^{(2)}$ are nearly the same ($\simeq 23.4$). In $I_2$ (same tonic cortical input of 3 Hz), ${\cal S}_{IP}^{(2)}$ becomes a little increased to 27.4 due to the inter-channel IP synaptic current from the channel 1 (with cortical input of 15 Hz). But, in $I_3$ (where the channel 2 receives the cortical input of 23 Hz), ${\cal S}_{DP}^{(2)}$ (= 7,562.5) is much larger than ${\cal S}_{IP}^{(2)}$ (= 2,139.1), mainly due to focused inhibition from D1 SPNs to the SNr via DP.

Then, the competition degree ${\cal C}_d^{(2)}$ of the channel 2 between DP and IP (given by the ratio of ${\cal S}_{DP}^{(2)}$ to ${\cal S}_{IP}^{(2)}$)
is shown in Fig.~\ref{fig:Ch2}(g). In $I_1$ (with the tonic cortical input of 3 Hz), ${\cal C}_d^{(2)} = 1.00$ (i.e., DP and IP are balanced), resulting in no action selection (i.e., BG gate to the thalamus is locked). In $I_2$ (where the channel 2 receives the same tonic cortical input of 3 Hz), ${\cal C}_d^{(2)}$ becomes decreased a little to 0.86, because of increased inter-channel IP synaptic current $I_{IP,inter}^{(2, 1)}(t)$ from the channel 1 (receiving the 15 Hz cortical input). Finally, in $I_3$ (with the cortical input of 23 Hz), ${\cal C}_d^{(2)} = 3.54$ (i.e., DP is 3.54 times stronger than IP). In this case, due to the
strong focused inhibition from the D1 SPNs to the SNr through DP (off-center effect), the BG gate to the thalamus becomes open, resulting in action selection; refer to Fig.~\ref{fig:ASS}(a3). In this time interval $I_3,$ action deselection of the channel 1 takes place [see Fig.~\ref{fig:ASS}(b)], because the increased inter-channel IP current $I_{IP,inter}^{(1, 2)}$ from the channel 2 suppresses the competing action in the channel 1 (on-surround effect, causing contrast enhancement to spotlight action selection in the channel 2). As a result, action switching occurs in $I_3$ from the channel 1 to the channel 2 with larger ${\cal C}_d^{(2)}$ (= 3.54).

\subsection{Quantitative Analysis for Action Selection in The Channel 3}
\label{subsec:QACh3}
Finally, we consider the case of the channel 3 with tonic cortical inputs of 3 Hz in all the time intervals, $I_1,$ $I_2,$ and $I_3,$ and make quantitative analysis for action selection. Figure \ref{fig:Ch3} shows quantitative analysis for action selection in the channel 3 through DP (green) and IP (red) synaptic currents into the channel 3. Detailed data, associated with the DP and IP synaptic currents, $I_{DP}^{(3)}(t)$ and $I_{IP}^{(3)}(t)$, are given in Figs.~\ref{fig:Ch3}(a1)-\ref{fig:Ch3}(d2). Interval-averaged DP and IP synaptic currents, $\overline{I_{DP}^{(3)}(t)}$ and $\overline{I_{IP}^{(3)}(t)}$, in each time interval, $I_1,$ $I_2,$ and $I_3,$ are shown in Figs.~\ref{fig:Ch3}(a1) and \ref{fig:Ch3}(a2), respectively.

Due to the tonic cortical inputs of 3 Hz, interval-averaged DP synaptic currents $\overline{I_{DP}^{(3)}(t)}$ (green) are very small (-23.4) in all the 3 time intervals, $I_1,$ $I_2,$ and $I_3,$ in contrast to the above cases of the channels 1 and 2. In $I_1,$ the interval-averaged IP synaptic current $\overline{I_{IP}^{(3)}(t)}$ (= 23.5; red) is nearly the same as the magnitude of $\overline{I_{DP}^{(3)}(t)}$. But, with increasing time interval,
$\overline{I_{IP}^{(3)}(t)}$ becomes increased to 27.6 in $I_2$ and 36.0 in $I_3,$ due to the increased inter-channel synaptic currents in Figs.~\ref{fig:Ch3} (d1) and \ref{fig:Ch3} (d2), $\overline{ I_{IP,inter}^{(3, 1)} }$ (= 754.0) from the channel 1 for $I_2$ and $\overline{ I_{IP,inter}^{(3, 2)} }$ (= 1,538.1) from the channel 2 for $I_3.$

Figure \ref{fig:Ch3}(e) shows population-averaged MFRs $\langle f_{{\rm SNr},i}^{(3)} \rangle$ of the SNr neurons in the channel 3 in $I_1,$ $I_2,$ and $I_3.$
In $I_1$ with tonic cortical input of 3 Hz, strengths of the DP and IP synaptic currents are nearly the same, and the SNr neurons fire actively with $\langle f_{{\rm SNr},i}^{(1)} \rangle=$ 25.8 Hz, leading to inhibition of the thalamus. With increasing the time intervals, $I_2$ and $I_3$, $\langle f_{{\rm SNr},i}^{(3)} \rangle$
increase to 28.4 Hz and 29.8 Hz, respectively, due to increase in the inter-channel IP synaptic currents, $I_{IP,inter}^{(3, 1)}$ in $I_2$ and $I_{IP,inter}^{(3, 2)}$ in $I_3$. Thus, the thalamus becomes more inhibited in $I_2$ and $I_3$.

Figures \ref{fig:Ch3}(f1) and \ref{fig:Ch3}(f2) show ${\cal S}_{DP}^{(3)}$ (strength of the DP synaptic current) and ${\cal S}_{IP}^{(3)}$ (strength of the IP synaptic current)  for the channel 3 in $I_1,$ $I_2,$ and $I_3$. In $I_1$, ${\cal S}_{DP}^{(3)}$ and ${\cal S}_{IP}^{(3)}$ are nearly the same ($\simeq 23.4$). With increasing the time intervals, $I_2$ and $I_3$, ${\cal S}_{IP}^{(3)}$ increases to 27.6 in $I_2$ and 36.0 in $I_3$, due to the increased inter-channel IP synaptic currents $I_{IP,inter}^{(3, 1)}$ in $I_2$ and $I_{IP,inter}^{(3, 2)}$ in $I_3$, while there is no essential change in ${\cal S}_{DP}^{(3)}$.
Thus, the competition degree ${\cal C}_d^{(3)}$ of the channel 3 in Fig.~\ref{fig:Ch3}(g) becomes decreased from 1.0 ($I_1$) to 0.85 ($I_2$) to 0.65 ($0.65$).
Consequently, in the case of the channel 3 receiving the tonic cortical inputs of 3 Hz in all the time intervals, no action selection occurs.

\subsection{Summary on The Off-center and On-surround Effect}
\label{subsec:OCOS}
We summarize our main results for the functions of DP and IP in terms of the competition degree, which is well shown in Fig.~\ref{fig:OCOS}.
Figures \ref{fig:OCOS}(a)-\ref{fig:OCOS}(b) show well the off-center and on-surround effect for action selection in $I_2$ (where the channel 1
receives cortical input of 15 Hz). Strong focused inhibition from the D1 SPNs is provided to the SNr via the DP synaptic current
$I_{DP}^{(1)}(t)$ (green), leading to suppress firing activity of the SNr (off-center effect). Consequently, the BG gate to the thalamus becomes opened, resulting in disinhibition of the thalamus. Hence, the major function of the DP is to suppress the population-averaged MFR $\langle f_{{\rm SNr},i}^{(1)} \rangle$  of the SNr, leading to action selection in the channel 1. In this case, the intra-channel IP synaptic current (red), $I_{IP,intra}^{({\rm STN},1)}$, from the STN in the channel 1 is also given to the SNr in the same channel. $I_{IP,intra}^{({\rm STN},1)}$ enhances $\langle f_{{\rm SNr},i}^{(1)} \rangle$  of the SNr, suppressing the desired action selection in the channel 1 (braking function). But, the effect of DP on the SNr is much larger than that of the intra-channel IP, resulting in the action selection in the channel 1.

We note diffusive excitation from the STN in the channel 1 to the SNr in the neighboring channels 2 and 3.
These inter-channel IP synaptic currents (red) from the STN in the channel 1, $I_{IP,inter}^{(2, 1)}$ and  $I_{IP,inter}^{(3, 1)}$, are
provided to the SNr in the channel 2 and  to the SNr in the channel 3, respectively. The inter-channel IP synaptic currents suppress competing actions in the channels 2 and 3 (on-surround effect), resulting in highlighting the desired action selection in the channel 1 via contrast enhancement.
In this way, the function of the inter-channel IP (suppressing the competing actions in the neighboring channels and spotlighting the desired action selection) is different from the braking function of the intra-channel IP to suppress the desired action selection in the same self-channel.

Figure \ref{fig:OCOS}(b) shows well quantitatively the off-center and on-surround effect in $I_2$ in terms of the competition degrees ${\cal C}_d^{(Ch)}$. In the previous time interval $I_1$ (gray), the competition degrees of all the 3 channels are nearly 1 (i.e., DP and IP are nearly balanced), leading to no action selection. But, in $I_2$ (black), ${\cal C}_d^{(1)}$ jumps to 5.68 (off-center effect) due to strong focused DP synaptic current to the SNr. On the other hand, ${\cal C}_d^{(2)}$ and ${\cal C}_d^{(3)}$ decrease to 0.86 and 0.85 (on-surround effect) due to the inter-channel IP synaptic currents from the STN in the channel 1 to the SNr in the channels 2 and 3, respectively. As a result of the on-surround effect, action selection in the channel 1 becomes highlighted due to contrast enhancement.

Next, we consider the case of the final time interval $I_3$ (where the channel 2 receives the cortical input of 23 Hz). We note that, in $I_3,$
action switching occurs from the channel 1 in $I_2$ to the channel 2. Due to the strong focused inhibitory DP synaptic current
$I_{DP}^{(2)}(t)$ (green) to the SNr in the channel 2 (off-center effect), action selection is made. In this case, as explained above, the intra-channel IP synaptic current $I_{IP,intra}^{({\rm STN},2)}$ serves a function of brake to suppress the desire action selection in the channel 2.

Because of the strong inter-channel IP synaptic current $I_{IP,inter}^{(1, 2)}$ from the STN in the channel 2 to the SNr in the channel 1 (on-surround effect), action deselection is made in the channel 1, resulting in action switching from the channel 1 to the channel 2. Likewise, another inter-channel IP synaptic current $I_{IP,inter}^{(3, 2)}$ to the SNr in the channel 3 suppresses the competing action in the channel 3. In this way, the inter-channel IPs serve the function of suppressing competing actions in the neighboring channels. Consequently, no interference occurs between the desired action in the channel 2 and the competing actions in the channels 3 and 1.

The off-center and on-surround effect in $I_3$ is well shown quantitatively in terms of the competition degrees ${\cal C}_d^{(Ch)}$ in Fig.~\ref{fig:OCOS}(d). We note that, in $I_3,$ ${\cal C}_d^{(2)}$ jumps to 3.54 (black) from 0.86 ($I_2$; gray) due to strong focused inhibitory DP synaptic current $I_{DP}^{(2)}(t)$ to the SNr in the channel 2 (off-center effect). Hence, action selection is made in the channel 2. On the other hand, ${\cal C}_d^{(1)}$ decreases from 5.68 ($I_2$) to 2.34 ($I_3$), due to the strong inter-channel IP synaptic current $I_{IP,inter}^{(1, 2)}$ to the SNr in the channel 1 (on-surround effect), and hence action deselection is made in the channel 1. In this way, action switching occurs from the channel 1 ($I_2$) to the channel 2 ($I_3$). In the case of the channel 3 (receiving only the tonic cortical input of 3 Hz), no action selection is also made in $I_3.$ Due to the inter-channel IP synaptic current $I_{IP,inter}^{(3, 2)}$ to the SNr in the channel 3 (on-surround effect), ${\cal C}_d^{(3)}$ decrease to 0.65 ($I_3$) from 0.85 ($I_2$).

In the above way, action selection is made in the channel with the largest competition degree ${\cal C}_d^{(Ch)}$ (where the population-averaged MFR
$\langle f_{{\rm SNr},i}^{(Ch)} \rangle$  of the SNr is the lowest) (off-center effect). Due to strong inter-channel IP synaptic currents to the neighboring channels, their competition degrees become decreased, leading to suppress competing actions in the neighboring channels (on-surround effect, causing contrast enhancement). Consequently, the on-surround effect leads to highlight the desired action selection, and no interference between desired action and competing actions occurs. In this way, functions of DP and IP for action selection (causing the off-center and on-surround effect) could be quantitatively made clear in terms of competition degrees ${\cal C}_d^{(Ch)}$.

\section{Summary and Discussion}
\label{sec:SUM}
In this paper, we are concerned about action selection performed by the BG in the SNN with 3 laterally interconnected channels.
A desired action is selected through strong focused inhibition from the D1 SPNs via the DP in a channel (off-center effect). There are two types of IPs because of diffusive excitation from the STN \cite{Parent1,Parent2,Parent3}. The intra-channel IP serves a function of brake to suppress the desired action in the corresponding channel. In contrast, the inter-channel IP to the SNr in the neighboring channels serves a function to suppress the competing actions, causing contrast enhancement (on-surround effect) \cite{Mink1,Mink2,Nambu,Triple,GPR1,GPR2,Hump1,Hump2,Hump3}. But, to the best of our knowledge, no quantitative analysis for the functions of the DP and the two intra- and inter-channel IPs was made.

Firing activity of the SNr (i.e., output nucleus of the BG) is well characterized in terms of their population-averaged MFR $\langle f_{{\rm SNr},i}^{(Ch)} \rangle$.
When $\langle f_{{\rm SNr},i}^{(Ch)} \rangle$ is high (low), the BG gate to the thalamus becomes locked (opened), leading to inhibition (disinhition) of the thalamus. In this way, $\langle f_{{\rm SNr},i}^{(Ch)} \rangle$ is a good indicator for the output activity of the BG, and hence it could also be used to determine a desired action selection \cite{Hump1,Spin}.
We note that firing activity (i.e. $\langle f_{{\rm SNr},i}^{(Ch)} \rangle$) of the SNr is determined via competition between the DP synaptic current and the IP synaptic current into the SNr. Their competition may be well characterized in terms of our recently-introduced competition degree $C_d^{(Ch)}$, given by the ratio of the strength of DP to the strength of IP \cite{KimPD}. In this way, $C_d^{(Ch)}$ plays a good role of indicator for the synaptic inputs into the SNr, in contrast to the output indicator, $\langle f_{{\rm SNr},i}^{(Ch)} \rangle$. Thus, relationship between $C_d^{(Ch)}$ and $\langle f_{{\rm SNr},i}^{(Ch)} \rangle$ could be regarded as the cause-and-effect. The larger $C_d^{(Ch)}$ is, the lower $\langle f_{{\rm SNr},i}^{(Ch)} \rangle$ becomes. In the channel with the lowest
$\langle f_{{\rm SNr},i}^{(Ch)} \rangle$, the BG gate to the thalamus is open (i.e., the thalamus becomes disinhibited, resulting in an action deselection). Consequently, a desired action may be selected in the channel with the largest ${\cal C}_d^{(Ch)}$ where $\langle f_{{\rm SNr},i}^{(Ch)} \rangle$ of the SNr neurons is the lowest. In the present work, we employed the competition degree  $C_d^{(Ch)}$ (input indicator) to determine a desired action selection, instead of the MFRs
$\langle f_{{\rm SNr},i}^{(Ch)} \rangle$ (output indicator).

Here, for a normal DA level ($\phi = 0.3$), we made quantitative analysis of functions of DP and IP for action selection by employing the competition degree ${\cal C}_d$ (characterizing competitive harmony between DP and IP) \cite{KimPD,KimHD}. We considered 3 competing channels.
For the channels 1 and 2, cortical inputs of 15 Hz and 23 Hz were applied from $t=1,000$ and 2,500 msec, respectively; for the channel 3, tonic cortical input of 3 Hz was applied. Desired action is selected in the channel with the largest ${\cal C}_d$. We have calculated the DP and the intra- and inter-channel IP synaptic currents into the SNr in each channel, and thus got the competition degree ${\cal C}_d^{(Ch)}$ of each channel to determine desired action. In the 1st time interval $I_1$ ($0 < t < 1,000$ msec), no action selection was made, because the competition degrees of the 3 channels (receiving tonic cortical inputs) were nearly the same ($\simeq 1.0$; DP and IP are balanced in each channel).

In the 2nd time interval $I_2$ ($1,000 < t < 2,500$ msec), desired action was selected in the channel 1 with the largest competition degree ${\cal C}_d^{(1)}$
(= 5.68; i.e., DP is 5.68 times stronger than IP), due to the focused inhibitory DP synaptic current to the SNr in the channel 1 (off-center effect). But, in the 3rd time interval $I_3$ ($2,500 < t < 5,000$ msec), action switching has been found to occur. The channel 2 had the largest competition degree ${\cal C}_d^{(2)}$ (= 3.54), and hence action selection was made in the channel 2. In $I_3$, the competition degree of the channel 1 was decreased to ${\cal C}_d^{(1)}=2.34$, due to strong inter-channel IP synaptic current from the channel 2 (on-surround effect), and hence action deselection was made. In contrast to the function of the inter-channel IP synaptic currents to suppress the competing action selections in the neighboring channels, in both $I_2$ and $I_3$ the intra-channel IP synaptic currents serve the function of brake to suppress the desired action selection in the corresponding channels. Through direct calculations of the DP and the intra- and inter-channel IP synaptic currents into the SNr in the 3 channels, functions of the DP and the intra- and the inter-channel IPs (causing the off-center and on-surround effect) have been quantitatively made clear in terms of the competition degree ${\cal C}_d^{(Ch)}$. Particularly, Fig.~\ref{fig:OCOS} has shown well the off-center and on-surround effect via the DP and the IP.

Finally, we discuss limitations of our present work and future works.
In the present work, we investigated action selection in the healthy state with harmony between DP and IP for the normal DA level ($\phi=0.3$).
But, through break-up of their harmony, pathological states such as Parkinson's disease (PD) and Huntington's disease (HD) occur \cite{KimPD,KimHD}.
Due to deficiency in DA (i.e. low DA level), the IP becomes stronger (i.e., it becomes over-active), leading to occurrence of PD (showing
hypokinetic movement disorder). On the other hand, because of degenerative genetic loss of D2 SPNs, the IP becomes weaker (i.e., it becomes under-active), resulting in occurrence of HD (exhibiting hyperkinetic movement disorder). Dysfunction in the BG circuitry (i.e., over- and under-active IP) was found to disrupt normal action selection process, which results in slower action switching, less efficiency at interference control of competing actions,
difficulty in inhibiting inappropriate responses, slower and inaccurate responses, and random switching between choices \cite{ASPD1,ASPD2,ASPD3,ASPD4,ASHD}.
As a future work, it would also be interesting to investigate quantitative analysis of action selection in the pathological states by employing our present approach for the normal state, based on the competition degree.

\section*{Acknowledgments}
This research was supported by the Basic Science Research Program through the National Research Foundation of Korea (NRF) funded by the Ministry of Education (Grant No. 20162007688).

\end{document}